\documentclass[fleqn,usenatbib,onecolumn]{mnras}

\usepackage{newtxtext,newtxmath}

\usepackage[T1]{fontenc}

\DeclareRobustCommand{\VAN}[3]{#2}
\let\VANthebibliography\thebibliography
\def\thebibliography{\DeclareRobustCommand{\VAN}[3]{##3}\VANthebibliography}

\usepackage{graphicx}	
\usepackage{amsmath}	
\usepackage{caption}
\usepackage{subcaption}
\usepackage{multicol}
\renewcommand{\bibpreamble}{\begin{multicols}{2}}
\renewcommand{\bibpostamble}{\end{multicols}}
\usepackage{enumitem}
\usepackage[dvipsnames]{xcolor}

\newcommand{\hi}{H\textsc{i}\ }
\newcommand{\hinospace}{\textrm{H\textsc{i}}}
\newcommand{\multidark}{\textsc{MultiDark}}
\newcommand{\deltadiff}{\delta\hspace{-0.2mm}}

\newcommand{\myparallel}{{\mkern4mu\vphantom{\perp}\vrule depth 1pt\mkern2.5mu\vrule depth 1pt\mkern4mu}}
\newcommand{\kFG}{k_{\!\myparallel}^{\mathrm{fg}}}

\usepackage{todonotes}

\newcommand{\secref}[1]{\hyperref[#1]{Section~\ref*{#1}}}
\newcommand{\appref}[1]{\hyperref[#1]{Appendix~\ref*{#1}}}

\makeatletter
\newcommand*\bigcdot{\mathpalette\bigcdot@{.5}}
\newcommand*\bigcdot@[2]{\mathbin{\vcenter{\hbox{\scalebox{#2}{$\m@th#1\bullet$}}}}}
\makeatother

\title[The \hi intensity mapping bispectrum]{The \hi intensity mapping bispectrum including observational effects}

\author[S. Cunnington et al.]{Steven Cunnington$^{1}$\thanks{E-mail: s.cunnington@qmul.ac.uk}, Catherine Watkinson$^{1}$, and Alkistis Pourtsidou$^{1,2}$
\\
$^{1}$School of Physics and Astronomy, Queen Mary University of London, Mile End Road, London E1 4NS, UK\\
$^{2}$Department of Physics \& Astronomy, University of the Western Cape, Cape Town 7535, South Africa
}

\date{Accepted XXX. Received YYY; in original form ZZZ}

\pubyear{2020}

\begin{document}
\label{firstpage}
\pagerange{\pageref{firstpage}--\pageref{lastpage}}
\maketitle

\begin{abstract}
The bispectrum is a 3-point statistic with the potential to provide additional information beyond power spectra analyses of survey datasets. Radio telescopes which broadly survey the 21cm emission from neutral hydrogen (\hinospace) are a promising way to probe LSS and in this work we present an investigation into the \hi intensity mapping (IM) bispectrum using simulations. We present a model of the redshift space \hi IM bispectrum including observational effects from the radio telescope beam and 21cm foreground contamination. We validate our modelling prescriptions with measurements from robust IM simulations, inclusive of these observational effects. Our foreground simulations include polarisation leakage, on which we use a Principal Component Analysis cleaning method. We also investigate the effects from a non-Gaussian beam including side-lobes. For a MeerKAT-like single-dish IM survey at $z=0.39$, we find that foreground removal causes a $8\%$ reduction in the equilateral bispectrum's signal-to-noise ratio $S/N$, whereas the beam reduces it by $62\%$. We find our models perform well, generally providing $\chi^2_\text{dof}\sim 1$, indicating a good fit to the data.  Whilst our focus is on post-reionisation, single-dish IM, our modelling of observational effects, especially foreground removal, can also be relevant to interferometers and reionisation studies.
\end{abstract}




\section{Introduction}

Observations of the Cosmic Microwave Background (CMB) reveal that fluctuations in our Universe's primordial density field are consistent with Gaussian fluctuations \citep{Aghanim:2018eyx}. Perfectly Gaussian random fields with zero mean are fully characterised by the two-point correlation function or its Fourier space equivalent, the power spectrum. As the Universe evolves, gravitational instability drives structure growth, a non-linear process, and hence causes departures from Gaussianity \citep{PeeblesBook,Bernardeau:2001qr}. Therefore, when probing large-scale structure (LSS) using late Universe observations, the information contained in the power spectrum is not a complete statistical description.

In order to extract information contained in the non-Gaussian components of LSS, higher order statistics can be used. Three-point correlation functions, and the Fourier counterpart referred to as the bispectrum, provide additional information not contained within the power spectrum. Measurements of clustering in galaxy catalogues using these 3-point statistics has been performed for decades \citep{PeeblesGoth1975,Groth:1977gj,FrySeldner1982,Jing:1998qs,Frieman:1999qj,Scoccimarro:2000sp,Verde:2001sf,Croton:2004hy,Jing:2003nb,Kulkarni:2007qu,Gaztanaga:2008sq,Marin:2010iv,Marin:2013bbb}. Correct interpretation of these measurements requires theoretical prescriptions to model non-linear matter and bias effects \citep{Fry1994,Angulo:2014tfa}. Furthermore, since observations of LSS are performed in redshift space, a correct treatment for the effect of redshift space distortions (RSD) on the bispectrum is required \citep{Hivon:1994qb,Matarrese:1997sk,Verde:1998zr,Heavens:1998es,Scoccimarro:1999ed,Scoccimarro:2000sn,Verde:2001sf,Sefusatti:2006pa}. With these techniques in place, measurements of the galaxy bispectrum have been performed in optical galaxy redshift surveys (e.g. \citet{Gil-Marin:2016wya,Pearson:2017wtw}), which have produced constraints on cosmological parameters. 
Next generation optical surveys such as DESI \citep{Aghamousa:2016zmz} and Euclid \citep{Blanchard:2019oqi} will soon be operational and will aim to to use the bispectrum to analyse the data they obtain. In combination with the power spectrum, this can help tighten the constraints on galaxy bias \citep{Yankelevich:2018uaz}.

A complementary approach to optical galaxy surveys for probing LSS is to use \hi intensity mapping (IM) \citep{Bharadwaj:2000av,Battye:2004re,Wyithe:2007rq,Chang:2007xk}. In the post-reionisation Universe, the vast majority of neutral hydrogen (\hinospace) is contained within galaxies, self-shielded from ionising radiation. This means that 21cm emission, caused from hyperfine transitions in \hinospace, will be a tracer of galaxies and thus the underlying matter density. \hi IM involves recording the unresolved, redshifted 21cm signals, in order to construct a 3-dimensional map of \hinospace. The advantage of this technique is that it has the potential to rapidly survey large cosmic-volumes covering a very wide redshift range, without being limited by high levels of shot noise. The 21cm signal will also be present out to very high redshifts and can therefore be used as a probe of the epoch of reionisation (EoR) \citep{Pritchard:2011xb, Patil:2017zqk} and even out to the cosmic dawn and the dark ages \citep{Bowman:2018yin}.

There are several observational challenges with the \hi IM method, for example the radio telescope's beam and 21cm foreground contamination. The signal captured by a radio telescope is received with some intensity pattern for each pointing. In some cases, especially single-dish IM \citep{batps}, this intensity pattern can be quite broad, with the full-width-half-maximum (FWHM) of the main beam (i.e. the central lobe) being over $1\,\text{deg}$ in size. The effect from this is to smooth density fluctuations transverse to the line-of-sight, suppressing information contained in small, perpendicular modes. Given that the observatories such as the Square Kilometre Array (SKA)\footnote{\href{https://www.skatelescope.org/}{skatelescope.org}} and its pathfinder MeerKAT will be reliant on the single-dish  method for its LSS science cases \citep{Bacon:2018dui,Wang:2020lkn}, this is an important observational effect to consider. 
Since with IM we aim to map the diffuse, unresolved \hi emission, observations become prone to accumulating foreground signals in the same frequency ranges as the redshifted \hinospace. These foreground contaminants are caused by numerous astrophysical processes such as cosmic-ray electrons accelerated by the Galactic magnetic field causing synchrotron radiation, or free-free emission caused by free electrons scattering off ions. Techniques exist to clean these foregrounds \citep{Liu:2011hh,Wolz:2013wna,Shaw:2014khi,Alonso:2014dhk,Cunnington:2020njn} but these inevitably also remove the \hi modes most degenerate with the foregrounds and can also leave some foreground residuals in the cleaned data, potentially biasing measurements.

Previous work has investigated the 21cm bispectrum at high redshift epochs, during the EoR and cosmic dawn \citep{Pillepich:2006fj,Shimabukuro:2015iqa,Yoshiura:2014ria,Watkinson:2017zbs,Majumdar:2017tdm,Bharadwaj:2020wkc,Mazumdar:2020bkm,Watkinson:2020zqg}. The bispectrum from post-reionisation \hi surveys has been studied in \citet{Sarkar:2019ojl}, who explored its real-space signatures with semi-analytical simulations, to probe the \hi bias. In reality however, \hi IM data will be recorded in redshift space and be subject to the effect of RSD.

Recent work has investigated the post-reionisation redshift space \hi intensity mapping bispectrum analytically (see e.g. \citet{Karagiannis:2019jjx, Durrer:2020orn,Jolicoeur:2020eup}), as well as higher-redshift studies with simulated signals \citep{Majumdar:2020kpt,Kamran:2020pkr}. However, a post-reionisation simulation-based analysis of the redshift space \hi IM bispectrum, including instrumental and foreground removal effects, is yet to be performed. 

In this work we investigate the prospects of performing bispectrum analyses using \hi IM. We include dedicated simulations of observational effects in the data, and we develop and test modelling prescriptions. The observational effects we consider come from RSD, the telescope beam, and 21cm foreground contamination and removal. We use an $N$-body simulation combined with a semi-analytical model, which is applied to generate gas masses for the galaxies and can be used to produce a brightness temperature for \hinospace. We emulate the observational effects in the simulated data, which allows for a comprehensive study of their signatures. Furthermore, we present modelling prescriptions for these effects and validate their performance using our simulations. We base our modelling on second order perturbation theory in redshift space, and also derive damping functions for the beam and foreground effects. The models we present should be beneficial in future analyses looking to detect the \hi IM bispectrum signal. In addition, the models should be applicable to analytical forecasts, making them more robust and reliable.

The paper is outlined as follows; in \secref{sec:Simulations} we introduce our simulated data, including the methods used to emulate the observational effects; in \secref{sec:BSmodelling} we outline the framework for modelling the \hi IM bispectrum in redshift space and modelling the observational effects, presenting validation tests throughout; finally, we summarise our main results and conclude in \secref{sec:Conclusion}.

\section{Simulated Data}\label{sec:Simulations}

Here we summarise our simulated data including the \hi signal with beam smoothing, as well as the 21cm foregrounds and their removal. We choose to tailor our simulations towards emulating a low-redshift \hi experiment, since this is consistent with current and forthcoming pathfinder surveys e.g. MeerKLASS (\citep{Santos:2017qgq}) and GBT  (\citep{Masui:2012zc,Wolz:2015lwa}). However, in principle, the modelling techniques we derive can be extended to interferometers and higher redshift studies.

\begin{figure}
    \centering
    \includegraphics[width=0.85\textwidth]{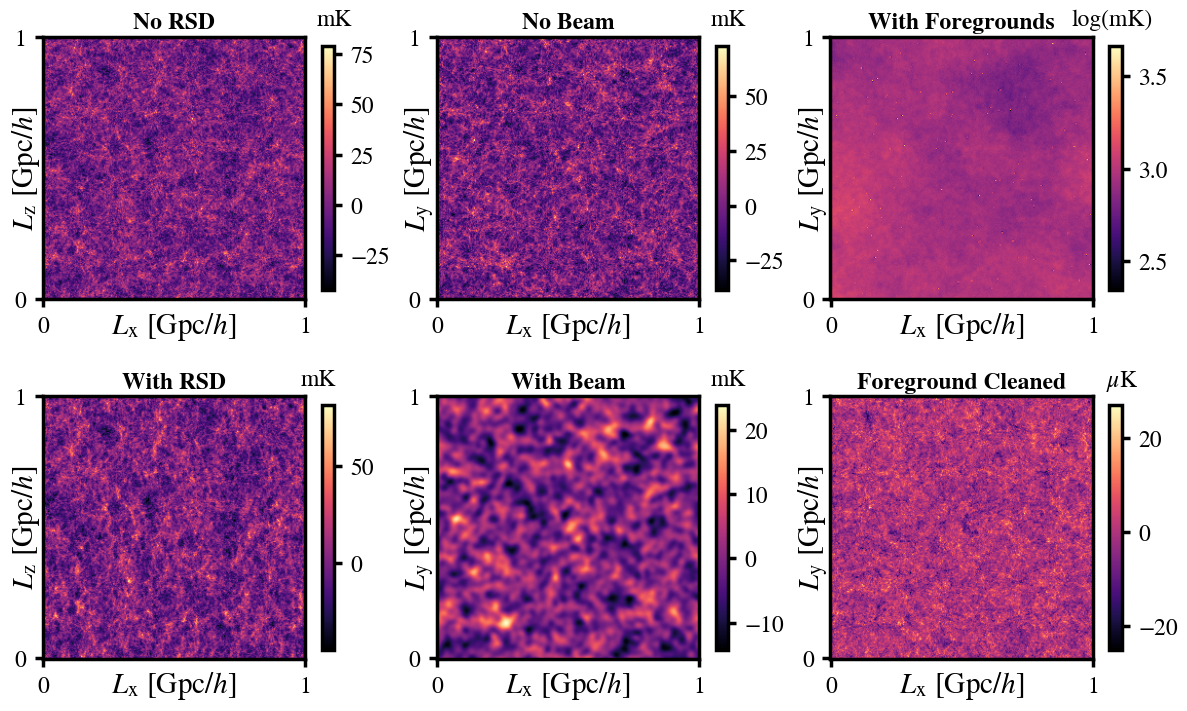}
    \caption{Maps of simulated data used in this study which have volume of $1\,(\text{Gpc}/h)^3$ and are at a central redshift of $z=0.39$. This is approximately equivalent to a sky survey of $3000\,\text{deg}^2$ and a redshift range of $\Delta z=0.4$, similar to the proposed MeerKLASS \hi IM survey using MeerKAT's L-band \citep{Santos:2017qgq,Pourtsidou:2017era}. 
    \textit{Left}-maps (averaged along the y-dimension) show the effects of RSD along the z-dimension (our chosen LoS direction). \textit{Middle}-maps (averaged along the z-dimension) show the effect of the radio telescope beam where the bottom map has been smoothed with a $R_\text{b}=10\,\text{Mpc}/h$ symmetric Gaussian kernel, acting on the dimensions perpendicular to the LoS (x and y) to emulate beam effects. \textit{Right}-maps (averaged along the z-dimension) demonstrate the effects from foregrounds where the top map is the full observed signal inclusive of 21cm foreground emission, and the bottom map has been cleaned by removing 10 modes using Principal Component Analysis (PCA).}
    \label{fig:Maps}
\end{figure}

Our underlying cosmological \hi simulation is based on the \textsc{MultiDark-Galaxies} $N$-body simulation \citep{Knebe:2017eei} with a semi-analytical application (\textsc{SAGE} \citep{Croton:2016etl}) to infer a \hi mass for each galaxy. To test our models, we select a low-redshift simulation snapshot at $z=0.39$ which is the approximate central redshift for a MeerKAT-like survey performed in the L-band ($899 < \nu < 1184\,\text{MHz}$, or equivalently $0.2<z<0.58$) \citep{Santos:2017qgq,Pourtsidou:2017era}. We outline the details for the simulations in \appref{app:CosmoHI} and also refer the reader to \citet{Cunnington:2020njn,Soares:2020zaq,Cunnington:2020mnn} where similar simulations were used. The final simulated data are over-temperature maps $\deltadiff T_\hinospace(\boldsymbol{x},z) = T_\hinospace(\boldsymbol{x},z) - \overline{T}_\hinospace(z)$, where $\overline{T}_\hinospace(z)$ represents the mean temperature of the field, and these are shown in \autoref{fig:Maps}. The maps on the left are averaged along the y-dimension and demonstrate the effects RSD have along the LoS (z-direction). The other maps are showing the effects of a telescope beam (middle) and foreground contamination (right), which we outline in the following sections. 

\subsection{Foreground Cleaning}

To investigate the effect foreground contamination has on the bispectrum, we simulated maps of 21cm foregrounds, added these onto the \hi IM, and then cleaned them using Principal Component Analysis (PCA). This will sufficiently emulate the damping of \hi power caused by foreground removal, and also produce any residual foreground contamination which is left in the data. We outline our approach to simulating the 21cm foreground maps, inclusive of polarisation leakage, in \appref{app:FGSims} and show a map of the foreground signal in \autoref{fig:Maps} (top-right). This demonstrates the dominance the foregrounds have over the \hinospace-only signal (see top-middle map for comparison, noting the log-scale used for the foregrounds).

Adding these foreground maps to the \hi creates foreground dominated simulated data. We then perform a PCA foreground clean, removing the first $N_\text{fg}$ principal components from the data's frequency-frequency covariance matrix. Since the foregrounds are dominant and highly correlated through frequency, i.e. along the LoS, this process removes the foreground signal leaving behind the cosmological \hi we are interested in. However, this method is imperfect and inevitably also removes \hi modes which are degenerate with the foregrounds, most typically large radial modes along the line-of-sight. 
Furthermore, not all the foreground is removed and residuals can be left in the data, especially where polarisation leakage is present, as is the case in our simulations. We the refer the reader to \citet{Cunnington:2020njn} for a detailed description of the PCA foreground cleaning method and its efficacy on \hi intensity maps contaminated with polarised foregrounds.

\autoref{fig:Maps} (bottom-right) shows a PCA cleaned intensity map with $N_\text{fg}=10$. Some differences between this and a foreground-free map (e.g. top-middle map) are immediately apparent. There is likely a large suppression of information due to removing 10 principal components from the data along with some residual foreground contamination. Furthermore, note the change of scale in the (bottom-right) colour bar relative to the other maps. Because the map has been averaged along the z-direction, and since a blind foreground clean, such as PCA, will remove each LoS's mean (as noted in \citet{Cunnington:2019lvb}), the range of fluctuations is restricted in this type of averaged map.

We note that future instruments should aim to have good control over calibration and if that is the case, less aggressive foreground cleaning would be required than what we simulate here. However, for this work, we opt for this conservative approach as a robust test on the limits of foreground contamination on the bispectrum.

\subsection{Telescope Beam}

The effect from the telescope beam is a smoothing to the temperature field in directions perpendicular to the LoS. A simple, and often sufficient, method to simulate these beam effects is to convolve the density field with a Gaussian kernel whose FWHM ($\theta_\text{FWHM}$) is chosen to match the model of the radio telescope one is trying to emulate. We can define this Gaussian smoothing kernel with \citep{batps}
\begin{equation}\label{eq:GaussianSpatialBeam}
    \mathcal{B}_\text{G}(\nu,\boldsymbol{s}_{\perp}) =  \exp\left[-4\ln2 \left(\frac{\boldsymbol{s}_{\perp}}{r(\nu)\,\theta_\text{FWHM}(\nu)} \right)^2 \right] = \exp\left[\frac{1}{2} \left(\frac{\boldsymbol{s}_{\perp}}{R_\text{b}} \right)^2 \right]\,,
\end{equation}
where $\boldsymbol{s}_{\perp} = \sqrt{\Delta x^2 + \Delta y^2}$ is the perpendicular spatial separation from the centre of the beam. $R_\text{b}= r(z)\,\sigma_\text{b}$ defines the physical size of the beam's central lobe in Mpc/$h$, where $\sigma_\text{b}=\theta_{\mathrm{FWHM}} /(2 \sqrt{2 \ln 2})$ represents the standard deviation of the Gaussian kernel in radians. $R_\text{b}$ is dependent on frequency through the comoving distance out to the the density fluctuations which changes with frequency ($r(\nu)$). It also has a further frequency dependence from the intrinsic beam size of the instrument, which is itself a function of frequency, generically given by $\theta_{\mathrm{FWHM}} \approx c/v D_{\mathrm{dish}}$, where $D_{\mathrm{dish}}$ is the diameter of the radio telescope dish.

Due to the added difficulty in modelling a frequency-dependent beam size, and also the complications it causes to foreground cleaning \citep{Matshawule:2020fjz}, it is common for data to be re-convolved to a common effective resolution (e.g. \citet{Masui:2012zc,Wolz:2015lwa}). The disadvantage of this procedure is the loss of information by smoothing the data to a larger resolution than the one caused by the telescope beam. For simplicity, we mostly assume this process has been employed which allows us to characterise different Gaussian beam cases by a single parameter and investigate the impact on the bispectrum by smoothing the \hi intensity maps with different values of $R_\text{b}$. The effect of a frequency-independent Gaussian beam is demonstrated by the middle maps of \autoref{fig:Maps} where a smoothing with $R_\text{b}=10\,\text{Mpc}/h$ has been performed on the lower map. For some context, a dish-size of $13.5\,\text{m}$ at $z=0.39$ (the dish size and effective redshift for a MeerKAT-like L-band survey) will result in a beam pattern with $R_\text{b}\sim12\,\text{Mpc}/h$. 

\begin{figure}
    \centering
    \includegraphics[width=0.85\textwidth]{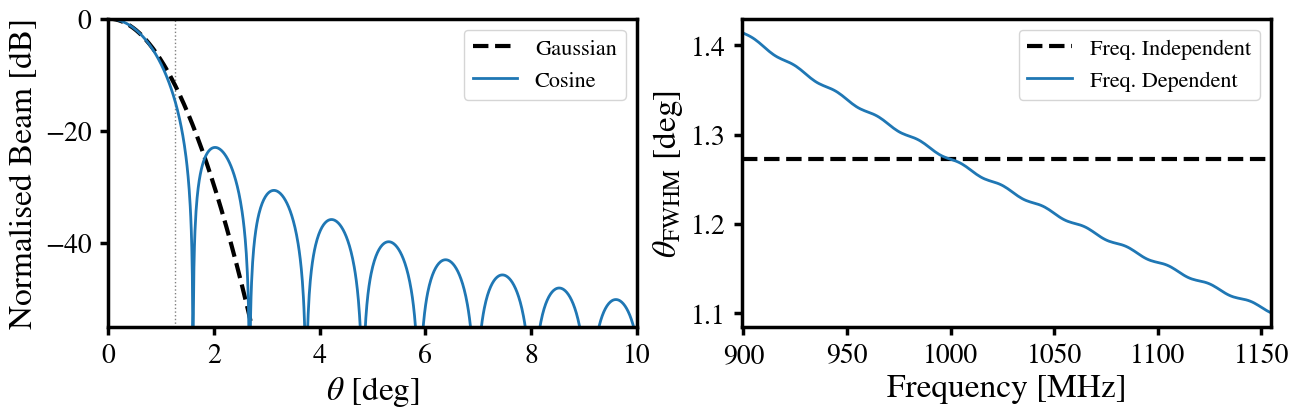}
    \caption{Different simulations for the telescope beam used in our analysis. \textit{Left}-panel shows the beam pattern for a generic Gaussian beam (\textit{black-dashed} line) and a more realistic Cosine beam (\textit{blue-solid} line) which includes multiple side-lobes as a function of angular distance from the beam centre $\theta$. The vertical \textit{grey-dotted} line marks the position of the FWHM for this example frequency which was chosen to be $1000\,\text{MHz}$. The \textit{right}-panel shows how the beam size given by $\theta_\text{FWHM}$ varies with frequency in a realistic beam simulation (\textit{blue-solid} line) against a constant frequency-independent beam (\textit{black-dashed} line).}
    \label{fig:BeamPattern}
\end{figure}

In order to investigate the effects from a more realistic beam, inclusive of side-lobes and with a complicated frequency dependence, we also simulate IM data where a cosine-tapered beam pattern has been applied, given by \citep{CondonBook} 
\begin{equation}\label{eq:CosineBeam}
    \mathcal{B}_{\mathrm{C}}(\nu,\boldsymbol{s}_\perp)=\left[\frac{\cos (1.189 \theta \pi / \theta_\text{FWHM}(\nu))}{1-4(1.189 \theta / \theta_\text{FWHM}(\nu))^{2}}\right]^{2}\,,
\end{equation}
where the beam size $\theta_\text{FWHM}$ is a function of frequency. For our simulations in Cartesian space, the angular separation $\theta$ from the centre of the beam can be given as $\theta = \boldsymbol{s}_{\perp} r(z)$. The side-lobes in this beam pattern are evident in the left panel of \autoref{fig:BeamPattern}, relative to the simple case of the Gaussian beam (black-dashed line) (\autoref{eq:GaussianSpatialBeam}). We have normalised the beam pattern such that it is 1 at the centre ($\theta=0$) but in its application in the simulations it is normalised such that its integral across whole sky region is 1. Noting the decibel scale of the y-axis, it is clear that the side-lobes are expected to be very small and will likely make negligible impact by eye on the \hi IM. However, it is still necessary to carefully test the departure from a purely Gaussian beam simulation. For example, side-lobes can cause issues in relation to foreground cleaning. This is due to the fact that the beam size, given by $\theta_\text{FWHM}$, changes with frequency. This can be a complicated, non-linear relationship and was investigated in \citet{Matshawule:2020fjz} where a \textit{ripple} model was provided. Based on this, we include a simplified version to introduce some frequency dependence, given by
\begin{equation}\label{eq:ripple}
    \theta_\text{FWHM}(\nu) = \frac{c}{\nu\,D_\text{dish}} + A \sin \left(\frac{2\pi\nu}{T}\right) \, ,
\end{equation}
where $A=0.1\,\text{arcmin}$ and $T=20\,\text{MHz}$. The second term in \autoref{eq:ripple} introduces a ripple into the frequency dependent relation of the beam size and can be seen in the right-panel of \autoref{fig:BeamPattern}. This frequency dependent beam pattern can cause issues for the foreground cleaning because point sources, or other foreground components not smooth in the angular directions, can oscillate in and out of a side-lobe's maxima due to the oscillating beam pattern shifting the side-lobe's position. This introduces a frequency structure to the foreground signal which can therefore be left in the data after a foreground clean which targets smooth, frequency coherent spectra. Whilst these structures will be minimal, due to the sub-dominant side-lobe power, they can still dominate the \hi signal. To ensure we include the potential for contamination from far-reaching side-lobes, we perform the beam convolution on a full-sky foreground map and outline the details for this in \appref{app:BeamSims}.

\section{\hi Intensity Mapping Bispectrum Modelling}\label{sec:BSmodelling}

Here we outline the framework for modelling the \hi IM bispectrum and present validation tests along the way with measurements from our simulated data. We base our modelling upon second order perturbation theory which is expected to hold only in the mildly non-linear regime \citep{Bernardeau:2001qr, Sefusatti:2006pa}. However, this should be sufficient for our primary purposes, which is modelling and testing observational effects on the \hi IM bispectrum.

The \hi bispectrum is defined by the 3-point function in Fourier space; 
\begin{equation}\label{eq:BSintro}
	\langle \delta T_\hinospace(\boldsymbol{k}_{\! 1}) \, \delta T_\hinospace(\boldsymbol{k}_{\!2}) \, \delta T_\hinospace(\boldsymbol{k}_{\!3}) \rangle = (2\pi)^3 B_\hinospace(\boldsymbol{k}_{\!1},\boldsymbol{k}_{\!2},\boldsymbol{k}_{\!3})\, \delta^\text{D}(\boldsymbol{k}_{\!1} + \boldsymbol{k}_{\!2} + \boldsymbol{k}_{\!3})\,,
\end{equation}
where $\delta T_\hinospace(\boldsymbol{k}) \equiv \int \text{d}^3 \boldsymbol{x}\,\delta T_\hinospace(\boldsymbol{x}) \exp(-i \boldsymbol{k} \bigcdot \boldsymbol{x})$ is the Fourier transform of the over-temperature field $\deltadiff T_\hinospace(\boldsymbol{x},z) = T_\hinospace(\boldsymbol{x},z) - \overline{T}_\hinospace(z)$. 
The Dirac delta function, $\delta^\text{D}(\boldsymbol{k}_{\!1} + \boldsymbol{k}_{\!2} + \boldsymbol{k}_{\!3})$, ensures the bispectrum is only defined for closed triangles of wavevectors. We begin by defining the \hi bispectrum in redshift space, and then we model the observational effects from \hi IM. We describe these by defining damping functions $D(\boldsymbol{k}_{\!i})$ which act on the bispectrum such that
\begin{equation}\label{eq:ObsModelBS}
    B^\hinospace_\text{obs}(\boldsymbol{k}_{\!1},\boldsymbol{k}_{\!2},\boldsymbol{k}_{\!3}) = B_\hinospace(\boldsymbol{k}_{\!1},\boldsymbol{k}_{\!2},\boldsymbol{k}_{\!3})\,D_\text{b}(\boldsymbol{k}_{\!1},\boldsymbol{k}_{\!2},\boldsymbol{k}_{\!3})\,D_\text{fg}(\boldsymbol{k}_{\!1},\boldsymbol{k}_{\!2},\boldsymbol{k}_{\!3})\,,
\end{equation}
where $B_\hinospace$ is the redshift space bispectrum which we will outline in \secref{sec:RSDmod}. $D_\text{b}$ and $D_\text{fg}$ are the damping functions for modelling the effects from the telescope beam and 21cm foreground contamination, derived in \secref{sec:BeamMod} and \secref{sec:FGMod}, respectively.

For measuring the bispectrum in our simulated data, we use the publicly available code \texttt{bifft}\footnote{\href{https://bitbucket.org/caw11/bifft/src/master/}{bitbucket.org/caw11/bifft/src/master}} \citep{Watkinson:2017zbs}. \texttt{bifft} exploits Fast-Fourier Transforms to enforce the Dirac-delta function of \autoref{eq:ObsModelBS} to drastically speed up the calculation of the bispectrum \citep{Scoccimarro:2015bla}. An operation that would naively be a series of nested loops through the dataset to find the bispectrum contribution of all triangles that conform to a given configuration, becomes one whose main overhead consists of six Fast-Fourier transforms and four loops through the dataset. In other words the code is fast, taking of order a few seconds per triangle configuration when run on a MacBookPro (2.3GHz i9 intel core, 16Gb RAM) for a datacube with $256^3$ pixels on a side. The method used by \texttt{bifft} is described and tested against a direct-sampling method in \citet{Watkinson:2017zbs} and \citet{Majumdar:2017tdm}. A succinct explanation of the code's inner workings is also provided in \citet{Watkinson:2020zqg}. We note that in this work we have also cross-checked \texttt{bifft} with another publicly available code, \texttt{Pylians}\footnote{\href{https://pylians3.readthedocs.io/en/master/Bk.html}{pylians3.readthedocs.io/en/master/Bk.html}} \citep{Villaescusa-Navarro:2018vsg}, and found agreement (\texttt{Pylians} applies the direct-sampling method).

\subsection{Redshift Space Bispectrum}\label{sec:RSDmod}

The matter density field $\delta$ is isotropic in real space. However, observations of \hi which trace the underlying density are conducted in redshift space and therefore a particular mode's measurement will now depend on its direction of alignment relative to the LoS i.e. $B(k_{\!1},k_{\!2},k_{\!3}) \rightarrow B(\boldsymbol{k}_{\!1},\boldsymbol{k}_{\!2},\boldsymbol{k}_{\!3})$. Therefore, any modelled \hi bispectra fitted to data will need to account for the anisotropies introduced by RSD. In this work we exclusively operate in a Cartesian space, thus the plane-parallel approximation is exactly valid and we can parameterise the alignment of modes to the LoS with $\mu_i = \boldsymbol{k}_{\!i}\bigcdot\hat{\text{z}}/k_i \equiv k_{i,\myparallel}/k_i$ \citep{Kaiser:1987qv}. Since the bispectrum is defined for closed triangles such that $\boldsymbol{k}_{\!3} = -(\boldsymbol{k}_{\!1}+\boldsymbol{k}_{\!2})$, the bispectrum in redshift space is a function of five variables. Following the formalism from \citet{Scoccimarro:1999ed}, we use three parameters which describe the shape of the triangle $k_{\!1},k_{\!2}$ and the angle $\theta$ between them i.e. $\cos\theta \equiv \hat{\boldsymbol{k}}_{\!1}\bigcdot\hat{\boldsymbol{k}}_{\!2}$. The other two variables describe the orientation of the triangle relative to the LoS; $\omega = \cos^{-1}(\mu_1)$ and the azimuthal angle $\phi$ about $\hat{\boldsymbol{k}}_{\!1}$. This provides the expressions;
\begin{equation}\label{eq:muDef}
    \mu_{1}=\mu=\cos \omega=\hat{\boldsymbol{k}}_{\!1} \bigcdot \hat{\text{z}}\, ,\, \quad \mu_{2}=\mu \cos \theta-\sqrt{\left(1-\mu^{2}\right)} \sin \theta \cos \phi\, ,\, \quad \mu_{3}=-\frac{k_{1}}{k_{3}} \mu-\frac{k_{2}}{k_{3}} \mu_{2}\,.
\end{equation}
On large scales in the linear regime, the \hi over-temperature field is given by $\delta T_\hinospace(\boldsymbol{k}) = \overline{T}_\hinospace b_\hinospace \delta(\boldsymbol{k})$ where $b_\hinospace$ represents the linear bias. The effect of measuring a Fourier component of this field in redshift space can be modelled as $\delta T_\hinospace(\boldsymbol{k}) \rightarrow \delta T_\hinospace^\text{s}(\boldsymbol{k}) = \overline{T}_\hinospace Z_1(\boldsymbol{k})\,\delta(\boldsymbol{k})$ \citep{Kaiser:1987qv}, where the superscript s denotes that the quantity is in redshift space. This is the only time we use the notation $\delta T_\hinospace^\text{s}({\boldsymbol{k}})$ to denote a quantity in redshift space. In all other cases we drop the supscript s for brevity. Unless clearly stated, we will always be working in redshift space. The factor $Z_1$ is often referred to as the Kaiser factor and is given by \citep{Kaiser:1987qv}
\begin{equation}\label{eq:Kaiser}
	Z_1(\boldsymbol{k}) = b_\hinospace + f\mu^2\,,
\end{equation}
where $f$ is the linear growth rate of structure, approximated by $f\simeq\Omega_\text{m}(z)^{0.55}$ \citep{Linder:2005in}. 

For future \hi IM observations, where we aim for precise measurements and constraints, it will be necessary to include non-linear effects in cosmological clustering statistics to avoid significant discrepancies in the determination of the \hi bias and other parameters \citep{Matarrese:1997sk,Mann:1997df,Castorina:2019zho}. Since using the bispectrum to break degeneracies and determine bias parameters is seen as one of its primary benefits, it is necessary to have accurate modelling prescriptions for it. From Eulerian perturbation theory, in which we assume a local, non-linear bias between the \hi over-density ($\delta_\hinospace$) and the underlying matter fluctuations ($\delta$), we can Taylor expand in $\delta$ \citep{Fry:1992vr};
\begin{equation}
	\delta_\hinospace = \sum_{i} \frac{b_{i}}{i !}\delta^i\,,
\end{equation}
and only retain terms up to $i=2$ (also ignoring $i=0$ which only contributes to $\boldsymbol{k}=0$), which leads to an expression for the biased \hi over-density field
\begin{equation}
	\delta_\hinospace \equiv \frac{\delta T_\hinospace}{\overline{T}_\hinospace} = b_1\delta+\frac{b_2}{2}\delta^2\,,
\end{equation}
where $b_1 \equiv b_\hinospace$ is the linear bias parameter and $b_2$ is the non-linear (second order) bias. Other studies have considered extensions to this which include compensation terms for non-local effects (e.g. \citet{Yankelevich:2018uaz}) which are due to gravitational evolution causing a non-local bias to develop in the halo distribution. It has been shown that for high precision cosmology, including these non-local bias terms is essential \citep{Chan:2012jj,Baldauf:2012hs} and omitting the corrections for these effects will certainly cause biased parameter estimation. However, in this work, where we aim to explore observational effects on the \hi IM bispectrum which should dominate over the non-local bias, we choose not to extend our model to incorporate this. We emphasise though that an exploration of the \hi IM bispectrum for precise cosmological parameter estimation would require this extension, something we leave for future work.

To describe the \hi bispectrum in redshift space, we apply the standard redshift space kernels (see \citet{Heavens:1998es,Scoccimarro:1999ed} for derivations) such that
\begin{equation}\label{eq:RSDBispec}
	B_\hinospace(\boldsymbol{k}_{\!1},\boldsymbol{k}_{\!2},\boldsymbol{k}_{\!3}) = 2\,\overline{T}_\hinospace^2\left[Z_1(\boldsymbol{k}_{\!1}) \, Z_1(\boldsymbol{k}_{\!2}) \, Z_2(\boldsymbol{k}_{\!1}, \boldsymbol{k}_{\!2}) \, P_\text{lin}(k_{\!1}) \, P_\text{lin}(k_{\!2}) + \text{cycl.}\right] D_\text{FoG}(\boldsymbol{k}_{\!1},\boldsymbol{k}_{\!2},\boldsymbol{k}_{\!3},\sigma_\text{B})\,,
\end{equation}
where cycl. represents cyclic permutations which run over all possible pairs of $\boldsymbol{k}_{\!1},\boldsymbol{k}_{\!2}$ and $\boldsymbol{k}_{\!3}$. $P_\text{lin}$ represents the real-space, linear matter power spectrum for which we use the the CLASS Boltzmann solver \citep{Lesgourgues:2011re,Blas:2011rf}. $Z_1$ is given in \autoref{eq:Kaiser} and $Z_2$ denotes the second-order kernel and is given by
\begin{equation}\label{eq:Z2}
	Z_2(\boldsymbol{k}_{\!i}, \boldsymbol{k}_{\!j}) = b_1 F_2(\boldsymbol{k}_{\!i}, \boldsymbol{k}_{\!j})+f \mu_{i\!j}^2 G_2(\boldsymbol{k}_{\!i}, \boldsymbol{k}_{\!j})+\frac{f \mu_{i\!j} k_{i\!j}}{2}\left[\frac{\mu_i}{k_i} Z_1(\boldsymbol{k}_{\!j})+\frac{\mu_j}{k_{\!j}} Z_1(\boldsymbol{k}_{\!i})\right] + \frac{b_2}{2} \, ,
\end{equation}
where $\boldsymbol{k}_{\!i\!j} = \boldsymbol{k}_{\!i} + \boldsymbol{k}_{\!j}$ and $\mu_{i\!j} = \boldsymbol{k}_{i\!j}\bigcdot\hat{\textbf{z}}/k_{\!i\!j}$. $F_2$ and $G_2$ denote the second-order kernels for the real-space density and velocity fields and are given by
\begin{equation}
	F_2(\boldsymbol{k}_{\!i}, \boldsymbol{k}_{\!j}) = \frac{5}{7} + \frac{m_{i\!j}}{2}\left(\frac{k_i}{k_{\!j}} + \frac{k_{\!j}}{k_i}\right) + \frac{2}{7} m_{i\!j}^2\,,
\end{equation}
\begin{equation}
	G_2(\boldsymbol{k}_{\!i}, \boldsymbol{k}_{\!j}) = \frac{3}{7} + \frac{m_{i\!j}}{2}\left(\frac{k_i}{k_{\!j}} + \frac{k_{\!j}}{k_i}\right)+\frac{4}{7} m_{i\!j}^2\,,
\end{equation}
where $m_{i\!j} = (\boldsymbol{k}_{\!i}\bigcdot \boldsymbol{k}_{\!j})/(k_i k_{\!j})$. The final term in \autoref{eq:RSDBispec}, $D_\text{FoG}$, is a phenomenological factor to address some non-linear RSD effects not sufficiently modelled by the redshift kernels alone. On smaller scales, internal motion inside virialized structures produces a radial smearing to the density field in redshift space, known as the Fingers-of-God (FoG) effect \citep{Jackson:2008yv}. It is common to include a term which describes the FoG \citep{Taruya:2010mx}, even when including higher order perturbation theory terms and should be seen as a phenomenological damping required to correct for non-linear effects \citep{Verde:1998zr,Gil-Marin:2014pva}. For our choice of model, this factor is given by \citep{Gil-Marin:2014sta}
\begin{equation}
	D_\text{FoG}(\boldsymbol{k}_{\!1},\boldsymbol{k}_{\!2},\boldsymbol{k}_{\!3},\sigma_\text{B}) = \left[1 + \frac{1}{2}\left(k_{\!1}^2 \mu_1^2 + k_{\!2}^2\mu_2^2 + k_{\!3}^2\mu_3^2\right)^2 \sigma_\text{B}^2\right]^{-2}\,.
\end{equation}
It is worth noting that it is necessary for galaxy surveys to also include modelling of shot noise caused by discreteness effects in their bispectra analyses. However, for \hi IM, where unresolved signal is integrated over, this should not be a limiting factor \citep{Spinelli:2019smg}. We therefore do not consider any treatment of shot noise in our analysis, making the assumption that this should be very low in \hi IM observations.

As a useful data-compression technique, and similar to the multipole expansion of the power spectrum into Legendre polynomials (see \citet{Cunnington:2020mnn, Soares:2020zaq} for applications to \hi IM), the dependence on the orientation of a triangle of wavevectors, parameterised by $\omega$ and $\phi$, can be decomposed into spherical harmonics \citep{Scoccimarro:1999ed,Scoccimarro:2015bla}
\begin{equation}
	B(\boldsymbol{k}_{\!1},\boldsymbol{k}_{\!2},\boldsymbol{k}_{\!3})=\sum_{\ell=0}^{\infty} \sum_{m=-\ell}^{\ell} B_{\ell m}(k_{\!1}, k_{\!2}, \theta) \, Y_{\ell m}(\omega, \phi)\,,
\end{equation}
where
\begin{equation}
	B_{\ell m}(k_{\!1}, k_{\!2}, k_{\!3}) = \int_{-1}^{+1} \int_{0}^{2\pi} B(\boldsymbol{k}_{\!1}, \boldsymbol{k}_{\!2}, \boldsymbol{k}_{\!3}) Y_{\ell m}^*(\theta,\phi)\, \text{d}\cos(\theta)\, \text{d}\phi\,.
\end{equation}
This shares the bispectrum signal between the different multipoles $B_{\ell m}$. To avoid working with the full multipole decomposition ($\ell,m)$, it is common to focus on the coefficients with $m=0$, referred to as the redshift-space multipoles and corresponds to averaging over $\phi$. In this case we can decompose the bispectrum with Legendre polynomials;
\begin{equation}
	B(\boldsymbol{k}_{\!1}, \boldsymbol{k}_{\!2}, \boldsymbol{k}_{\!3}) = \sum_{\ell=0}^{\infty} B_\ell(k_{\!1}, k_{\!2}, \theta) \mathcal{L}_{\ell}(\mu)\,.
\end{equation}
In this analysis we focus solely on the monopole, and we refer the interested reader to \citet{Yankelevich:2018uaz} for the higher order multipoles description. The monopole ($\ell=0$) with $\mathcal{L}_0=1$, equates to an averaging over $\mu$ so that the bispectrum monopole is given by
\begin{equation}
    B_0(\boldsymbol{k}_{\!1},\boldsymbol{k}_{\!2},\boldsymbol{k}_{\!3})=\frac{1}{4 \pi} \int_{-1}^{+1} \text{d}\mu \int_{0}^{2 \pi} \text{d}\phi \,B(\boldsymbol{k}_{\!1},\boldsymbol{k}_{\!2},\boldsymbol{k}_{\!3},\omega, \phi)\,.
\end{equation}
\begin{figure}
\begin{minipage}{.48\textwidth}
    \centering
    \includegraphics[width=\textwidth]{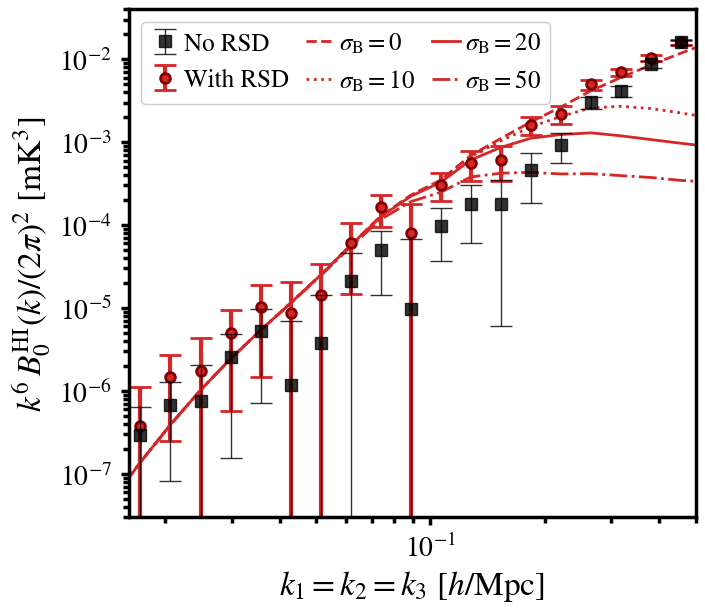}
    \captionof{figure}{Equilateral redshift space \hi bispectrum monopole from simulated \hi intensity maps. We model the RSD case with the \textit{red-dashed} line and find good agreement with the data (\textit{red-circular} points). The other line-styles show models with differing $\sigma_\text{B}$. For comparison, we also show the bispectrum for the simulation in real space, i.e. without RSD (\textit{black-square} points).}
    \label{fig:NoRSD}
\end{minipage}
\hfill
\begin{minipage}{.48\textwidth}
    \centering
    \includegraphics[width=\textwidth]{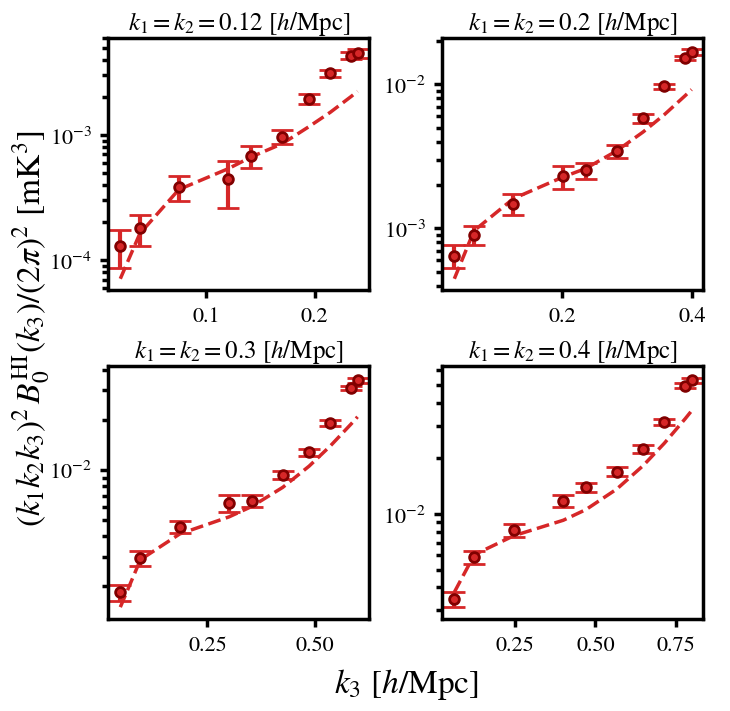}
    \captionof{figure}{\hi bispectrum monopole for isosceles configurations, for four different fixed sizes of $k_1=k_2$ (given in panel titles) as a function of a varying $k_3$. Overlaid as \textit{dashed} line is our model inclusive of RSD effects with $\sigma_\text{B}=0$.}
    \label{fig:RSDisos}
\end{minipage}
    \centering
    \includegraphics[width=1\textwidth]{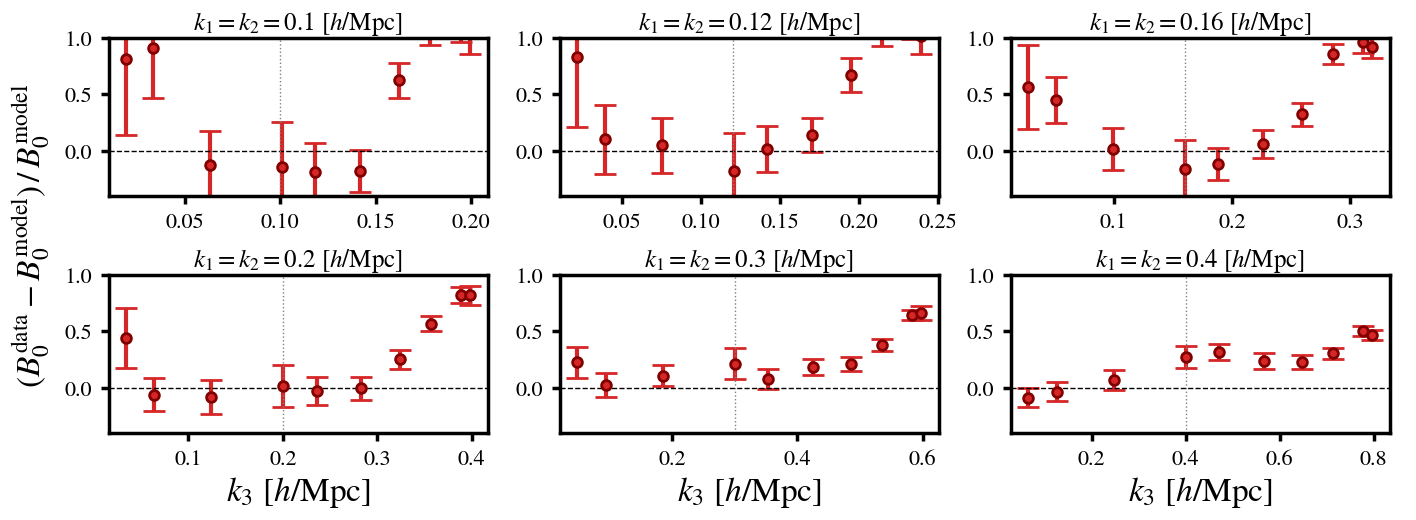}
    \caption{Residual comparison between data and model for redshift space bispectrum monopole for \hi IM for different isosceles configurations. Vertical \textit{grey-dotted} line marks the point where $k_3 = k_1,k_2$ i.e. the equilateral configuration.}
    \label{fig:NonLinearLimit}
\end{figure}
We begin by investigating an equilateral triangle configuration of the bispectrum, a special case with $k_{\!1}=k_{\!2}=k_{\!3}$. 
In \autoref{fig:NoRSD} we plot the measured equilateral bispectrum monopole for the simulated \multidark\ \hi IM, omitting for now any beam, foreground, or thermal noise effects, which we introduce in the following sections. We show the dimensionless\footnote{This normalisation is commonly referred to as dimensionless in the literature since spatial dimensions have been normalised out. However, for radio IM, the bispectrum will still have units of $\text{mK}^3$.} bispectrum, $k^6B(k)/(2\pi)^2$, and stick to the dimensionless convention in all subsequent plots. The red-circular data points represent \textit{redshift} space measurements but we also plot the \textit{real} space measurements (black-squares) to demonstrate the effects RSD have on the bispectrum and motivate their modelling in order to avoid biased (incorrect) results. In order to obtain error-bars in \autoref{fig:NoRSD} (and subsequent plots) we employ a jackknifing technique \citep{Norberg:2008tg} using 64 jackknife regions to compute the covariance matrix and use the diagonal elements as our error estimates. We found the contribution to the covariance from the off-diagonal elements is minimal, with the exception of the small-$k$ bins (see \appref{app:Covariance} and \autoref{fig:Covariance}). In any case, the uncorrelated error assumption we make is reasonable for our purposes. We find errors increase with decreasing $k$ due to cosmic-variance limitations as one would expect. We note that a realistic experiment would contain some thermal noise and would therefore slightly increase errors on smaller scales. However, as we demonstrate later in \secref{sec:Noise}, purely thermal noise should have little impact on the \hi IM bispectrum. It is the results with RSD which we fit the redshift space \hi bispectrum model to (red-dashed line). We use known input fiducial parameters $\overline{T}_\hinospace = 0.0743\,\text{mK}$ and $f=0.714$. This leaves three remaining free parameters, namely $\{b_1,b_2,\sigma_\text{B}\}$. Best-fit analyses using these parameters have been performed in the optical galaxy surveys literature, for example in \citet{Gil-Marin:2014sta}. At the low redshifts we consider, these works have found that the modelling breaks down already at $k>0.15 \, h/{\rm Mpc}$. 
Since our main goal in this paper is to quantify and model the observational effects related to \hi IM observations, with the beam effects dominating at the non-linear regime, we do not attempt to perform a best-fit analysis. Instead, we find sensible values by eye and keep them fixed throughout this work, concentrating on the modelling of the beam and foreground removal effects. These are $\{b_1,b_2,\sigma_\text{B}\} = \{1.5,2.3,0\}$. 
We demonstrate the effect of changing $\sigma_\text{B}$ in \autoref{fig:NoRSD} by plotting some non-zero values (note that \citet{Gil-Marin:2014sta} find $\sigma_\text{B} \sim 10$ for their tracers, but trusting the model only up to $k_{\rm max} = 0.17 \, h/{\rm Mpc}$). As can be seen this damps the bispectrum at high-$k$, as expected. However, we find this makes agreement worse with our data, which include highly non-linear scales up to $k_{\rm max} = 0.5 \, h/{\rm Mpc}$. Our simulations should encapsulate some FoG effects but we expect a smaller contribution in \hi IM relative to galaxy surveys where detected galaxies are in general more exclusively hosted by the highest mass haloes. The FoG effects can indeed be seen still by comparing the RSD simulation results to the no-RSD case (black-squares) at high-$k$. Here, we see the RSD data begin to lower in amplitude relative to the no-RSD results, thus likely evidence of FoG. Therefore, the lack of agreement at high-$k$ with a FoG model can be explained by a failing of the model at these non-linear scales. We will return to the discussion on non-linear effects and modelling in \secref{sec:NonLinear}. With the caveats discussed above in mind, we find a good agreement between data and model, with a $\chi_\text{dof}^2\sim 1$. 

\autoref{fig:RSDisos} shows measured bispectra for different isosceles triangle configurations. We fix $k_1$ and $k_2$ to four different values as shown in the panel titles, then plot results for a varying $k_3$. Again, we overlay our model (dashed line) and see reasonable agreement. The isosceles models also show a tendency to under-predict at high-$k$ and have perhaps greater discrepancies than the equilateral results of \autoref{fig:NoRSD}. It is also not easy to identify a fixed scale at which all models breakdown but we discuss this next in \secref{sec:NonLinear}. We note that we have not applied any corrections for aliasing effects, which could in principle be causing some discrepancies at high-$k$. However, we performed tests to investigate this and found no evidence it is causing noticeable effects at the scales we are interested in. We discuss this further in \appref{app:Aliasing}.

\subsubsection{Non-Linear Effects}\label{sec:NonLinear}

As we have already mentioned, our simulated data are at a low redshift, $z=0.39$, in order to emulate current pathfinder surveys. Sufficiently modelling non-linear scales at these redshifts is difficult, and current state-of-the-art models struggle to accurately model the biased redshift space bispectra above $k_{\max}\sim 0.15\,h/\text{Mpc}$ \citep{Gil-Marin:2014sta,Lazanu:2015rta,Yankelevich:2018uaz} for the purposes of precision cosmology. Developing an accurate model of a \hi IM bisepctrum well into non-linear scales that cannot be treated perturbatively is beyond the aims of the paper. With this considered, our approach is performing as one would reasonably expect from previous work. 

In general, we expect non-linear effects to become more important at high-$k$ and there will therefore be some maximum scale which we can sufficiently model up to. \autoref{fig:NoRSD} shows that even with our mildest assumption of FoG ($\sigma_\text{B}=10$) we begin to see a divergence between data and model at $k\gtrsim 0.2\,h/\text{Mpc}$. The $\sigma_\text{B}=0$ value we have chosen by eye still provides a sensible fit, likely because of the interplay between FoG and other non-linear effects.

However, we only find this to be the case for the equilateral configuration. When we analyse different isosceles cases, we see more conclusive discrepancies at high-$k$. \autoref{fig:NonLinearLimit} shows the agreement between data and model for different isosceles configurations. In general, we achieve a high signal-to-noise ratio and good agreement for $k\gtrsim0.03\,h/\text{Mpc}$ (below this, the error for cosmic variance is large). Eventually though, agreement starts to worsen at higher-$k$ due to non-linear effects, likely due to only using a leading-order (tree-level) bispectrum model. We find that agreement begins to deteriorate at different scales depending on the configuration. In the $k_1=k_2=0.4\,h/\text{Mpc}$ case, we see good agreement up to $k_3\sim 0.6\,h/\text{Mpc}$. But discrepancies begin at much lower $k_3$ for lower $k_1$, $k_2$ values. The vertical grey dotted line in \autoref{fig:NonLinearLimit} marks the equilateral triangle point, i.e. where the configuration moves from squeezed ($k_1,k_2 < k_3$) to squashed ($k_1,k_2>k_3$). In general we find that model agreement begins to suffer when the configuration moves into the squashed regime. 

We highlight here that there may also be some limitations from our simulation which has a limited mass resolution of $\sim10^9h^{-1}$M$_\odot$ per dark matter particle, as detailed in \appref{app:CosmoHI}. A more conclusive investigation of the non-linear modelling of \hi  bispectra (and power spectra) would be very valuable for 21cm precision cosmology, but this would require highly sophisticated, ideally hydrodynamical, simulations with improved mass resolutions such as IllustrisTNG (see investigation in \citet{Villaescusa-Navarro:2018vsg}). Since the velocity dispersion is larger for higher mass haloes, it is possible there will be some differences in the FoG effect between \hi IM and a spectroscopic galaxy survey. Galaxy surveys are generally populated by galaxies in the highest mass haloes, whereas \hi IM detects signal down to the lowest mass host haloes. Therefore, whilst \hi IM should still have a greater FoG effect than the underlying dark matter, since haloes with masses less than $10^8\,h^{-1}M_\odot$ should stop hosting \hi \citep{Villaescusa-Navarro:2018vsg,Modi:2019ewx}, in principle the impact from FoG should still be lower in \hi IM than conventional galaxy surveys. However, since the primary aim of this work is to investigate the observational effects more unique to \hi intensity mapping, namely the telescope beam and foreground contamination, we leave a more detailed analysis of non-linear effects in the \hi field for future work.

\subsection{Modelling the
Beam}\label{sec:BeamMod}

The effect from the telescope beam is to smooth the density field in all directions perpendicular to the LoS and therefore its effect on a Fourier component of the \hi over-temperature field can be modelled as $\delta T_\hinospace(\boldsymbol{k}) \rightarrow \delta T^\text{sm}_\hinospace(\boldsymbol{k}) = \mathcal{B}_\text{b}\delta T_\hinospace(\boldsymbol{k})$, where $\delta T_\hinospace^\text{sm}$ denotes a smoothed quantity and $\mathcal{B}_\text{b}$ represents the beam function (see \appref{app:BeamSims} for more details). For the case of a Gaussian frequency-independent beam, i.e. one whose FWHM does not vary in size along the LoS and has constant size given by the physical scale $R_\text{b}$, we can Fourier transform the Gaussian beam function in \autoref{eq:GaussianSpatialBeam} to get the beam function
\begin{equation}
    \mathcal{B}_\text{b}(\boldsymbol{k}) = \exp \left[-\frac12k_{\!\perp}^{2} R_\mathrm{b}^2\right] = \exp \left[-\frac{k^2}{2}(1-\mu)^{2} R_\mathrm{b}^2\right]\,.
\end{equation}
This demonstrates that the beam will damp large $k_\perp$ modes. We can model this effect on the bispectrum by considering the combined contributions from smoothed modes, which provides the damping term required for \autoref{eq:ObsModelBS}, and is given by
\begin{equation}
    D_\text{b}(\boldsymbol{k}_{\!1},\boldsymbol{k}_{\!2},\boldsymbol{k}_{\!3}) = \mathcal{B}_\text{b}(\boldsymbol{k}_{\!1})\,\mathcal{B}_\text{b}(\boldsymbol{k}_{\!2})\,\mathcal{B}_\text{b}(\boldsymbol{k}_{\!3}) = \exp \left\{-\left[\frac{k_{\!1}^{2}}{2} \left(1-\mu_1^{2}\right) + \frac{k_{\!2}^{2}}{2} \left(1-\mu_2^{2}\right) + \frac{k_{\!3}^{2}}{2} \left(1-\mu_3^{2}\right)\right]R_{\mathrm{b}}^{2}\right\}\,.
\end{equation}
Thus the \hi bispectrum monopole, with observational effects from the telescope beam, can be modelled as
\begin{equation}\label{eq:BeamModel}
   B^\hinospace_0(\boldsymbol{k}_{\!1},\boldsymbol{k}_{\!2},\boldsymbol{k}_{\!3}) = \int^{+1}_{-1}\int^{2\pi}_{0} B_\hinospace(\boldsymbol{k}_{\!1},\boldsymbol{k}_{\!2},\boldsymbol{k}_{\!3})\, \exp\left\{-\left[\frac{k_{\!1}^{2}}{2} \left(1-\mu_1^{2}\right) + \frac{k_{\!2}^{2}}{2} \left(1-\mu_2^{2}\right) + \frac{k_{\!3}^{2}}{2} \left(1-\mu_3^{2}\right)\right]R_{\mathrm{b}}^{2}\right\} \text{d}\mu\,\text{d}\phi\,,
\end{equation}
where $B_\hinospace$ is the redshift space bispectrum in \autoref{eq:RSDBispec} and expressions for $\mu_1,\mu_2,\mu_3$ can be found in \autoref{eq:muDef}.

\begin{figure}
\begin{minipage}{.48\textwidth}
    \centering
    \includegraphics[width=\textwidth]{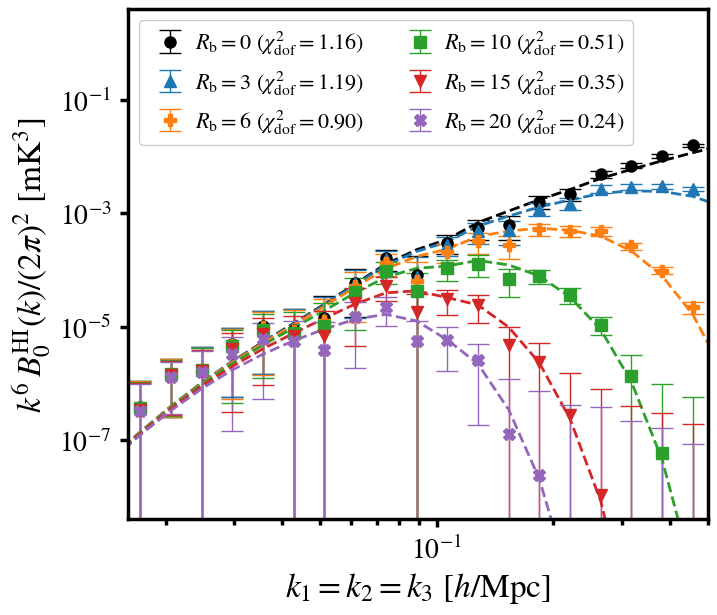}
    \captionof{figure}{Effects on the equilateral \hi bispectrum monopole from different beam sizes denoted by $R_\text{b}$. All cases are for a simple Gaussian, frequency independent beam. Models using \autoref{eq:BeamModel}, are overlaid as \textit{dashed} lines. We show the reduced $\chi^2_\text{dof}$ measurements for each beam case in the legend, which demonstrate a good agreement between model and data in most cases.}
    \label{fig:BeamModel}
\end{minipage}%
\hfill
\begin{minipage}{.48\textwidth}
    \centering
    \includegraphics[width=\textwidth]{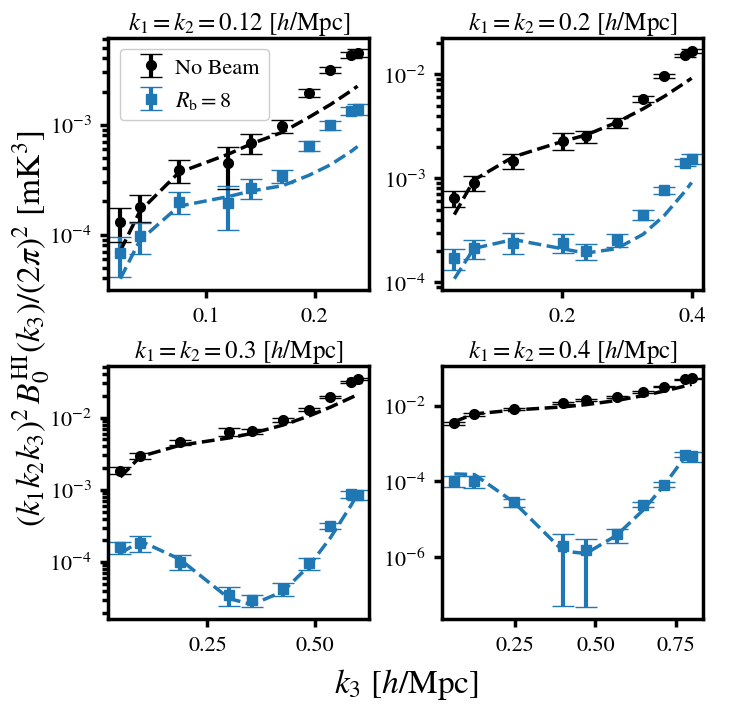}
    \captionof{figure}{Effects of a $R_\text{b}=8\,\text{Mpc}/h$ Gaussian frequency-independent beam on different isosceles configurations for the \hi IM bispectrum monopole (\textit{blue-squares}). Included for comparison, are the no beam ($R_\text{b}=0$) case (\textit{black-circles}) results. \textit{Dashed}-lines are the models using \autoref{eq:BeamModel}.}
    \label{fig:Beamisos}
\end{minipage}
    \centering
    \includegraphics[width=1\textwidth]{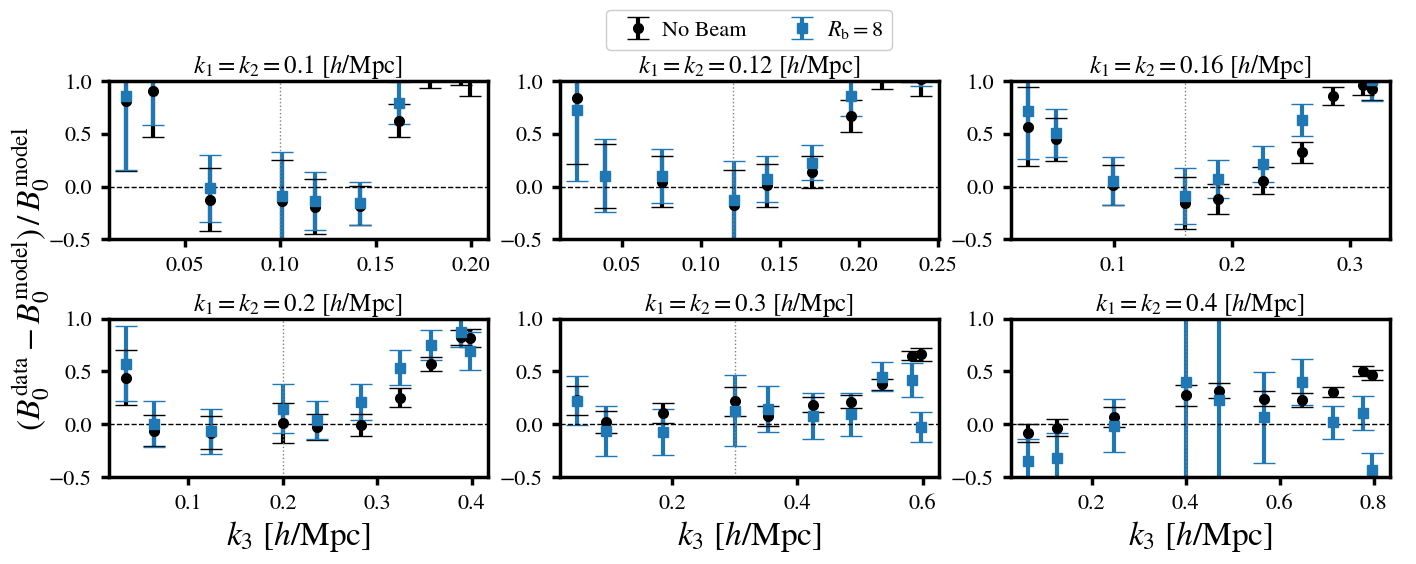}
    \caption{Residual comparison between no beam and $R_\text{b}=8\,\text{Mpc}/h$ for the \hi IM bispectrum monopole. We show the agreement between data and model for different isosceles configurations. Vertical \textit{grey-dotted} line marks the point where $k_3 = k_1,k_2$ i.e. the equilateral configuration.}
    \label{fig:BeamModvData}
\end{figure}

We begin by showing this model applied to our simulated data measurements for the equilateral case, where we know, from the previous results in \autoref{fig:NoRSD}, that the model without beam effects was working sufficiently well. In \autoref{fig:BeamModel} we show the results for a range of increasing $R_\text{b}$ values which represents an increasing physical beam size. The impact from an increasing beam is seen in the results which show that a higher $R_\text{b}$ amounts to more damping on the bispectrum at high-$k$, as expected, since the contributions from high-$k_{\!\perp}$ are being restricted. 

We find our model agrees well with the simulated data results across all beam sizes we test for the equilateral monopole and show the $\chi^2_\text{dof}$ results in the legend of \autoref{fig:BeamModel}. For modest beam sizes we see a good agreement with $\chi^2_\text{dof}\sim 1$ but find this steadily decreases for high-values of $R_\text{b}$, which can be indicative of over-fitting. This could mean that we are over-estimating the errors for the highly-smoothed cases which would require revising our jackknifing routine. Alternatively, this could mean the errors estimates are reasonable and we are fitting data consistent with zero-signal at high-$k$ in the extreme levels of high-smoothing. Neither of these explanations would suggest a poor performing beam model and we thus conclude that this is a sufficient model for a Gaussian beam. 

We can see further evidence for a well performing model in the isosceles configurations from \autoref{fig:Beamisos} where the damping from the beam introduces a less trivial distortions to the shape of the bispectra. In these cases we use one beam size of $R_\text{b}=8\,\text{Mpc}/h$ and compare this to the no beam ($R_\text{b}=0$) case. Despite the less trivial distortions to the bisepctra, our modelling seems consistent in all cases with measurements from the simulated data. Again, we see some general discrepancies at higher-$k$, this is the best demonstrated by \autoref{fig:BeamModvData} where we show a direct data model comparison for different isosceles cases. We again attribute the high-$k$ discrepancies to non-linear effects. Since the telescope beam is a smoothing of the field, it is plausible to expect some alleviation of non-linear effects at high-$k$, however, this is only a smoothing of modes perpendicular to the line-of-sight, and hence non-linear effects can still dominate radially.

These results are only for the simple case of a Gaussian beam. We will investigate the impact from more complex beam patterns in \secref{sec:NonGaussBeam}.

\subsection{Modelling Foreground Contamination}\label{sec:FGMod}

As discussed in \secref{sec:Simulations}, several methods exist for removing dominant foregrounds from \hi intensity maps but these are imperfect and cause some contamination and damping to the power spectrum and bispectrum. The most notable of these effects is the damping of large modes along the LoS, which are the modes most degenerate with the foregrounds. This will cause a reduction in contribution from small-$k_{\!\myparallel}$ modes which we look to model here. To test this model we employ a PCA method, the most commonly used approach to foreground cleaning, which removes the first $N_\text{fg}$ principal components from the frequency-frequency covariance matrix.

\begin{figure}
\begin{minipage}{.48\textwidth}
    \centering
    \includegraphics[width=\textwidth]{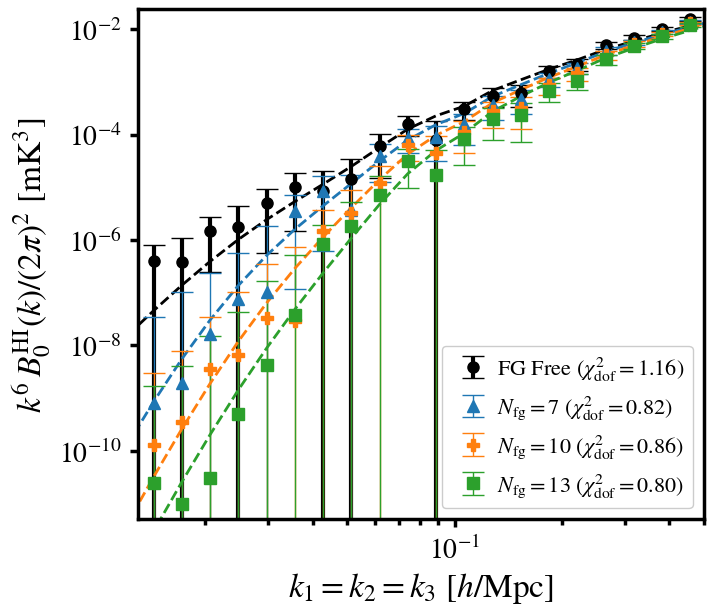}
    \captionof{figure}{Effects on the equilateral \hi bispectrum monopole from different levels of foreground cleaning. The \textit{black-circle} points represent \hi intensity maps with no foregrounds and therefore no foreground clean is required. For the other data, $N_\text{fg}$ denotes the number of components removed in a PCA clean. The \textit{dashed}-lines are the models using \autoref{eq:FGmodel} with $k_\parallel^\text{fg} = 1.2\times10^{-2},2.3\times10^{-2}, 3.5\times10^{-2}\,h/\text{Mpc}$ respectively. The $\chi^2_\text{dof}$ measurements for each case are shown in the legend, which demonstrate a good agreement between model and data.}
    \label{fig:FGModel}
\end{minipage}%
\hfill
\begin{minipage}{.48\textwidth}
    \centering
    \includegraphics[width=\textwidth]{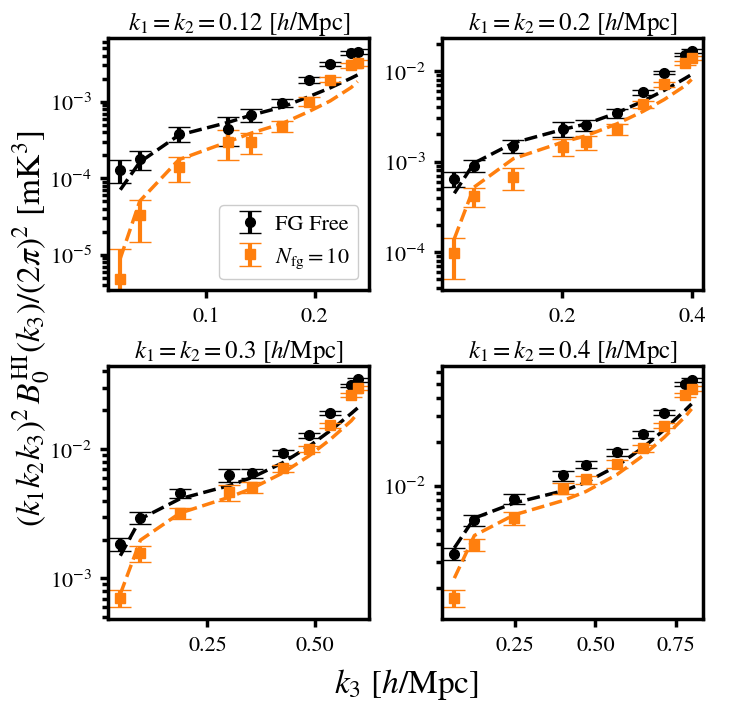}
    \captionof{figure}{Effects of a $N_\text{fg}=10$ PCA foreground clean (\textit{orange-squares}), on different isosceles configurations for the \hi IM bispectrum monopole. Included for comparison, are the foreground-free (\textit{black-circles}) results.}
    \label{fig:FGisos}
\end{minipage}
    \centering
    \includegraphics[width=1\textwidth]{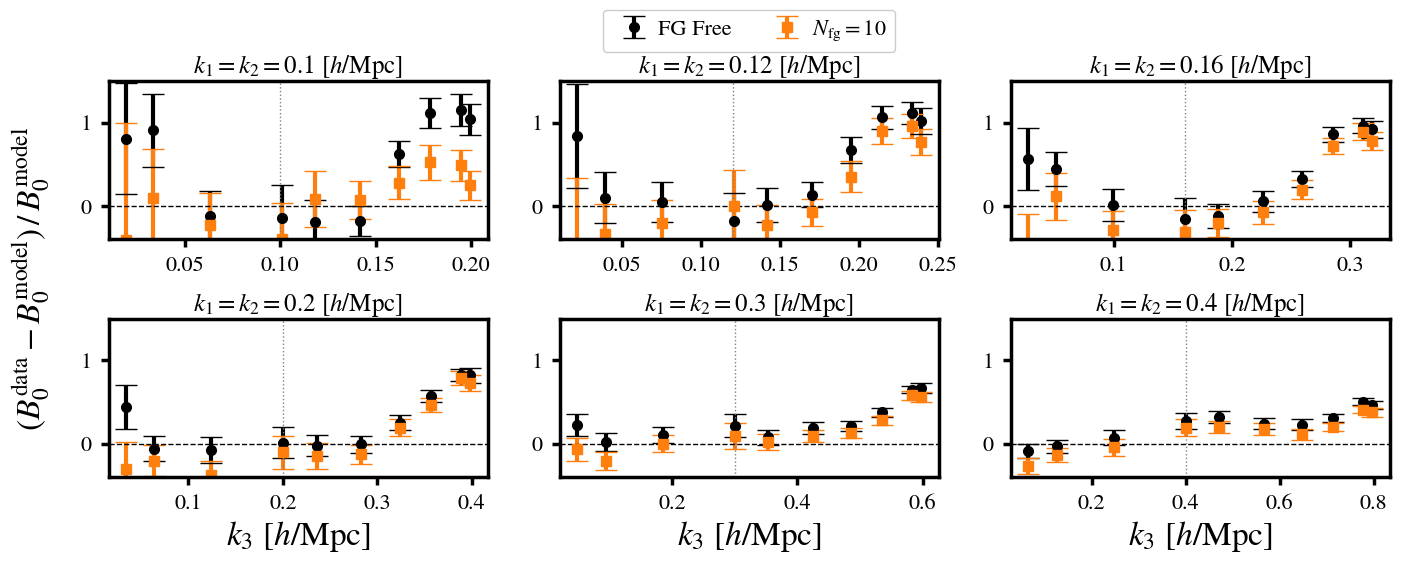}
    \caption{Residual comparison between foreground-free and foreground contaminated \hi IM bispectrum monopole. We show the agreement between data and model for different isosceles configurations. Vertical \textit{grey-dotted} line marks the point where $k_3 = k_1,k_2$ i.e. the equilateral configuration.}
    \label{fig:FGModvData}
\end{figure}

We can model the impact from this clean on a Fourier component of the \hi over-temperature field by considering $\delta T_\hinospace(\boldsymbol{k}) \rightarrow \delta T^\text{clean}_\hinospace(\boldsymbol{k}) = \mathcal{B}_\text{fg}\delta T_\hinospace(\boldsymbol{k})$ where $\delta T_\hinospace^\text{clean}$ denotes a Fourier component from a foreground cleaned IM and $\mathcal{B}_\text{fg}$ represents the foreground cleaning damping function \citep{Bernal:2019jdo}. To capture the main impact from foreground cleaning, we thus require $\mathcal{B}_\text{fg}$ to be some function which progressively damps the contribution to the signal from small-$k_{\!\myparallel}$ modes. There are various ways to perform this and we choose to adapt and extend the modelling used \citet{Soares:2020zaq}. For the power spectrum, the foreground damping function was given as
\begin{equation}
    \mathcal{B}_\text{fg}(\boldsymbol{k}) = 1-\exp \left[-\left(\frac{k_{\!\myparallel}}{k_{\!\myparallel}^{\mathrm{fg}}}\right)^2\right] = 1-\exp \left[-\left(\frac{k \mu}{\kFG}\right)^2\right]\,,
\end{equation}
where $\kFG$ is a free parameter governing the extent of information loss due to foreground cleaning. A higher $\kFG$, would mean more severe damping to modes from a more aggressive clean. We can model this effect on the bispectrum by considering the combined contributions from foreground damped modes, which provides the damping term required for \autoref{eq:ObsModelBS}, and is given by
\begin{equation}
    D_\text{fg}(\boldsymbol{k}_{\!1},\boldsymbol{k}_{\!2},\boldsymbol{k}_{\!3}) = \mathcal{B}_\text{fg}(\boldsymbol{k}_{\!1})\,\mathcal{B}_\text{fg}(\boldsymbol{k}_{\!2})\,\mathcal{B}_\text{fg}(\boldsymbol{k}_{\!3}) = \Bigg\{1-\exp \left[-\left(\frac{k_{\!1} \mu_1}{\kFG}\right)^2\right]\Bigg\}\Bigg\{1-\exp \left[-\left(\frac{k_{\!2} \mu_2}{\kFG}\right)^2\right]\Bigg\}\Bigg\{1-\exp \left[-\left(\frac{k_{\!3} \mu_3}{\kFG}\right)^2\right]\Bigg\}\,.
\end{equation}
Thus the \hi bispectrum monopole, with damping caused by a foreground clean, can be modelled as 
\begin{multline}\label{eq:FGmodel}
    B^\hinospace_0(\boldsymbol{k}_{\!1},\boldsymbol{k}_{\!2},\boldsymbol{k}_{\!3}) = \int^{+1}_{-1}\int^{2\pi}_{0} B_\hinospace(\boldsymbol{k}_{\!1},\boldsymbol{k}_{\!2},\boldsymbol{k}_{\!3})\, \Bigg\{1 - \exp \left[-\left(\frac{k_{\!1}\mu_1}{\kFG}\right)^2\right] - \exp \left[-\left(\frac{k_{\!2}\mu_2}{\kFG}\right)^2\right] - \exp \left[-\left(\frac{k_{\!3}\mu_3}{\kFG}\right)^2\right] + \exp \left[-\frac{k^2_{\!1}\mu^2_1 + k^2_{\!2}\mu^2_2}{\left(\kFG\right)^2}\right]\\ + \exp \left[-\frac{k^2_{\!1}\mu^2_1 + k^2_{\!3}\mu^2_3}{\left(\kFG\right)^2}\right] + \exp \left[-\frac{k^2_{\!2}\mu^2_2 + k^2_{\!3}\mu^2_3}{\left(\kFG\right)^2}\right] - \exp \left[-\frac{k^2_{\!1}\mu^2_1 + k^2_{\!2}\mu^2_2 + k^2_{\!3}\mu^2_3}{\left(\kFG\right)^2}\right]\Bigg\}\, \text{d}\mu\,\text{d}\phi\, ,
\end{multline}
where $B_\hinospace$ is the redshift space bispectrum in \autoref{eq:RSDBispec} and expressions for $\mu_1,\mu_2,\mu_3$ can be found in \autoref{eq:muDef}.

For investigating the impact from foregrounds we have removed the telescope beam from the simulation to avoid compounding two strong observational effects (although we test this combination in \secref{sec:NonGaussBeam}). We still include RSD in the simulations, hence our use of the redshift space bispectrum in \autoref{eq:FGmodel}. We begin by presenting results in the equilateral configuration and \autoref{fig:FGModel} shows the measured bispectrum for our simulated \hi IM inclusive of the foreground contamination (outlined in \appref{app:FGSims}), then foreground cleaned to different levels, parameterised by $N_\text{fg}$, which are the number of principal components removed from the frequency-frequency covariance. A higher $N_\text{fg}$ will remove more foreground contaminant but damp the \hi signal more drastically and this is what we see in \autoref{fig:FGModel} at small-$k$. Unlike the modelling for the Gaussian beam, where we knew the exact value for the parameter $R_\text{b}$ needed to model the results, the foreground clean model is more phenomenological and requires fitting the free parameter $\kFG$ to match results. This represents a flexible way to account for foreground cleaning effects since $\kFG$ parameter can be treated as a nuisance parameter and marginalised over when constraining cosmological parameters, as demonstrated in \citet{Cunnington:2020wdu}.

To model the three cases of $N_\text{fg} = 7,10,13$ we use fitted values of $\kFG=1.2\times10^{-2},2.3\times10^{-2}, 3.5\times10^{-2}\,h/\text{Mpc}$ respectively. These provide good reduced $\chi^2_\text{dof}$ results as shown in the legend of \autoref{fig:FGModel}. We see a similar trend to that seen in the beam results of \autoref{fig:BeamModel} where the $\chi^2_\text{dof}$ are decreasing as the bispectrum is damped more severely. This is most likely indicating that we are over-fitting the severely damped modes (in this case at low-$k$) that are fairly consistent with zero, but have a large error from the jackknifing process. 

We then explore some isosceles configurations in \autoref{fig:FGisos} where a $N_\text{fg}=10$ foreground clean was performed on the simulated data. We also provide the foreground-free results for comparison and the effects from the foreground clean are fairly intuitive with, in general, more damping to the bispectrum at smaller-$k$ in the foreground cleaned results. It is interesting to note that this is more evident in the changing $k_1 = k_2$ values, i.e. the top-left panel appears to show more damping than the bottom-right. Whereas the damping appears slightly more uniform across the range of $k_3$. Again we show the respective models as dashed lines, which are following these features measured in the data and in all cases are showing good agreement. The direct data and model comparison in \autoref{fig:FGModvData} demonstrates this nicely. Again we see some discrepancies between data and model, generally at high-$k$ from non-linear effects, but these discrepancies are consistent between the foreground-free modelling and the foreground cleaned one. In other words, we see no evidence that these discrepancies are exacerbated in the foreground contamination cases and thus conclude that the model is performing well.

Our main focus in this work regarding foreground effects has concerned signal attenuating, occurring typically on larger modes. However, a foreground clean will also inevitably leave some residual foreground in the data which could contaminate a bispectrum measurement across all modes by causing an additive bias. Whilst simulations seem to suggest that this should be a minor impact, but still relevant for precision cosmology \citep{Cunnington:2020njn}, it is clear that systematics are a big problem in real \hi IM data sets \citep{Switzer:2013ewa}. So far we have relied on cross-correlations with optical surveys to bypass the large systematics currently contained within pathfinder IM observations \citep{Masui:2012zc}, but it remains unclear how much of this systematic contribution is coming from residual foregrounds.

Finally, we should note that an alternative approach for addressing the damping from foreground cleaning is to employ a foreground transfer function (see \citet{Switzer:2015ria,Cunnington:2020njn} for details) which is a data-driven approach using mocks, injected with the real foreground and systematics contaminated data, then cleaned. The impact on the mock clustering statistics is then used to construct the transfer function which is applied to the real data to reverse the \hi signal loss effects from the foreground clean. This technique has been used for \hi IM power spectra measurements with pathfinder surveys \citep{Masui:2012zc,Switzer:2013ewa,Anderson:2017ert, Wolz:2021ofa}. However, using this in the context of a bispectrum measurement would be more cumbersome and computationally expensive. Importantly, previous works have shown that foreground removal effects can be degenerate with cosmological parameters \citep{Cunnington:2020wdu,Soares:2020zaq}. Therefore, for precision cosmology the transfer function approach needs to be studied further.

\subsection{Noise Contributions}\label{sec:Noise}

Unlike galaxy surveys, shot noise should not be a limiting factor for a radio telescope conducting a \hi IM survey \citep{Battye:2004re, Chang:2007xk}, a claim which has been supported by simulations \citep{Villaescusa-Navarro:2018vsg,Spinelli:2019smg}. Instead, the main source of noise comes from thermal motion of electrons inside the electronics of the instrument which produce Gaussian-like fluctuating currents, with a mean current of zero but a non-zero rms. The consequence from this is a component of white-noise contained in the maps. From the radiometer equation, the rms of the thermal noise contained in time-ordered data for an instrument with system temperature $T_\text{sys}$, with frequency and time resolution $\delta\nu$ and $\delta t$, will be given by $T_\text{sys}/\sqrt{\delta\nu\,\delta t}$ \citep{Wilson2009}. At map level this will create a field of white noise added into the data, with rms $\sigma_\text{N}$. In the case of the power spectrum this produces an additive component;
$P_\hinospace \rightarrow P_\hinospace + P_\text{N}$ where
$P_\text{N} = \sigma_\text{N}^2/V_\text{cell}$. However, since the fluctuations in this thermal noise are Gaussian, the thermal noise bispectrum should be zero and introduce no additive component to the modelled \hi IM bispectrum. However, statistical fluctuations of the noise do introduce an error contribution to the bispectrum. This was investigated and concluded in the context of EoR observations in \citet{Yoshiura:2014ria}.

We investigated this in our simulations by adding on to our \hi IM, a Gaussian field fluctuating around zero, with a rms of $\sigma_\text{N}$. As expected we found no additive bias to the bispectrum from these tests with a range of $\sigma_\text{N}$ but did find an increase in the errors obtained from our jackknifing procedure. This is demonstrated in \autoref{fig:SNR} (top-panel) where we plot the signal-to-noise ($S/N$) i.e. the ratio between the amplitude and the error ($\delta\!B_0$) on the bispectrum, for the \hi IM monopole as a function of $k$ for the equilateral configuration. The increase in errors, and thus a reduction in $S/N$, from thermal noise is very marginal at low-$k$ but more significant at higher-$k$. However, we have used extremely high values of $\sigma_\text{N}$ to demonstrate this. Indeed, a value of $\sigma_\text{N}=0.5\,\text{mK}$ is essentially unrealistic, since even calibration data from MeerKAT with a relatively low number of observational hours, should be able to achieve levels of $\sigma_\text{N}=0.2\,\text{mK}$ after some averaging \citep{Wang:2020lkn}. A level of $\sigma_\text{N}=0.05\,\text{mK}$ should be achievable with near term IM experiments, and this level of noise seems to have very little impact on the $S/N$. We include this level of noise in the bottom-panel (green-squares) of \autoref{fig:SNR} along with the foregrounds and beam, whose impact on the $S/N$ we discuss in the following section. Here we still see, that even in the presence of effects from the beam and a foreground clean, the noise causes no major change to the $S/N$, just a slight decrease at high-$k$.

It is thus encouraging to conclude that for a generic \hi IM survey, bispectrum measurements should be quite immune to the thermal noise from the instrument, causing only a mild reduction in $S/N$ and no thermal noise bias. Whilst the Gaussian assumption regarding thermal noise is a reasonable one, \hi IM is likely to have additional noise-like contributions from systematic effects such as RFI and 1/$f$ processes \citep{Harper:2018ncl,Li:2020bcr}. These could provide non-Gaussian contributions and thus sufficient calibration would be required to avoid biasing the bispectrum. This presents the possibility for using the bispectrum to characterise systematics in pathfinder \hi IM surveys, but we leave an exploration of this for future work. 

\subsection{Non-Gaussian Beam with Side-Lobes}\label{sec:NonGaussBeam}

The beam results we presented in \secref{sec:BeamMod} assumed a Gaussian beam with a $R_\text{b}$ that does not change with frequency. In reality the beam pattern is more complex, contains side-lobes, and will change in size as a function of frequency. For \secref{sec:BeamMod}, we therefore assumed that a perfect re-convolution had been performed to completely neutralise these complexities. This represents an unrealistic assumption and we will now investigate the consequences of relaxing it.

\begin{figure}
\begin{minipage}{.48\textwidth}
    \centering
    \includegraphics[width=\textwidth]{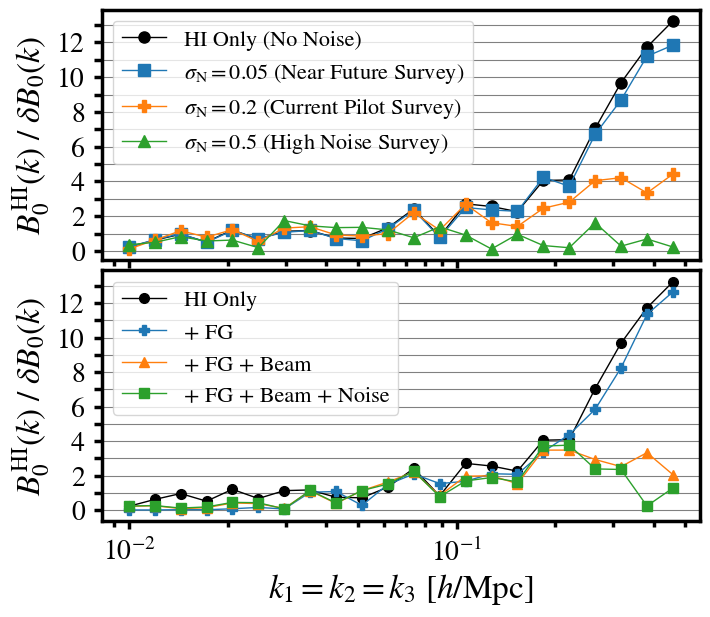}
    \captionof{figure}{The \hi IM bispectrum monopole $S/N$, given by $B_0^\hinospace/\delta 
    \!B_0$. \textit{Top-panel} shows an increasing level of thermal noise contained in the IM. $\sigma_\text{N}$ is the rms of the random Gaussian fluctuations, which closely model a thermal noise contribution. We explain the chosen $\sigma_\text{N}$ levels in the text. \textit{Bottom-panel} shows the impact on $S/N$ from including the foreground clean, then the beam, then the noise.}
    \label{fig:SNR}
\end{minipage}
\hfill
\begin{minipage}{.48\textwidth}
    \centering
    \includegraphics[width=\textwidth]{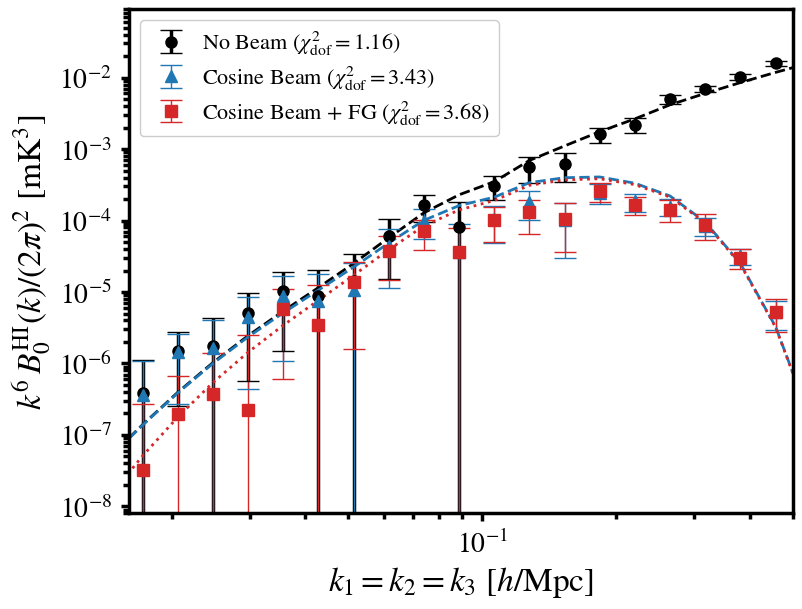}
    \captionof{figure}{Results from a more realistic beam simulation, which has a cosine beam pattern with multiple side-lobes and a frequency-dependent beam size (see \secref{sec:Simulations} and \autoref{fig:BeamPattern} for details). The case with foregrounds (red-squares) does not include polarisation leakage and has had a $N_\text{fg}=4$ PCA foreground clean.}
    \label{fig:CosineBeamModel}
\end{minipage}
\hfill
\end{figure}

A non-Gaussian beam with side-lobe structure could introduce some troubling effects into a statistical measurement of non-Gaussianity. We introduced the cosine beam pattern with side-lobes in \autoref{fig:BeamPattern} and we now investigate the results from a bispectrum measurement of the \hi IM simulations with this beam applied. \autoref{fig:CosineBeamModel} shows the bispectrum monopole results from the IM with a cosine beam for the equilateral configuration (blue triangles). We attempt to model this with the same Gaussian beam model as before (\autoref{eq:BeamModel}) using a value of $R_\text{b}=6.75\,\text{Mpc}/h$ fitted by eye (blue dashed line). The agreement looks quite reasonable, however, the reduced $\chi^2_\text{dof}$ statistic for this fit is shown in the legend, and it is clear that this is indicating a poorer fit relative to the no-beam case (black-circles), and the Gaussian beam cases from \autoref{fig:BeamModel}. This is perhaps expected since we are using the same Gaussian beam model, but now on IM data with a non-Gaussian beam. Furthermore, we have not made any corrections for the frequency-dependence in the beam given by \autoref{eq:ripple} and demonstrated in \autoref{fig:BeamPattern} (right-panel).

We also show the impact on $S/N$ from the cosine beam pattern in \autoref{fig:SNR} (bottom-panel) relative to the beam-free case. We see $S/N$ reduced by $50\%$ at $k\sim0.25\,h/\text{Mpc}$ when introducing the beam, although we found a similar reduction in $S/N$ is also present when using a Gaussian beam. This large reduction is unsurprising since the beam is damping the high-$k$ modes, which otherwise have very strong $S/N$. Comparing this to the impact from introducing the foregrounds (blue-crosses in \autoref{fig:SNR}), this also reduces the $S/N$ by $50\%$ at $k\sim0.03\,h/\text{Mpc}$. However, we can look at the overall reduction to $S/N$ by summing in quadrature each bin's contribution. We find that without the beam or foreground contamination we achieve $S/N = 22.9$, but we stress that this only considering the equilateral configuration. A complete and accurate forecast would include contributions from all triangle orientations and a comprehensive analysis of the covariance properties. With this in mind and just focusing on the equilateral configuration, we find foregrounds reduce the overall $S/N$ by $7.9\%$ whereas the beam reduces the overall $S/N$ by $61.9\%$. This suggests that the beam has a significantly larger impact on the detection of the bispectrum than the foreground contamination. However, it is likely that in real data analyses the beam will be more understood and thus easier to model. Therefore, the foregrounds could cause more problems for parameter estimation since they have the potential to bias results if not sufficiently understood.

The amplitudes of the side-lobes in the cosine beam pattern (\autoref{fig:BeamPattern}) are very small relative to the central lobe, and it is perhaps unsurprising that they only mildly degrade the agreement with the bispectrum model. However, combining this more realistic beam pattern with foreground contamination can potentially create drastic systematic effects. As discussed in \secref{sec:Simulations}, this is due to beam size oscillating with frequency, and thus point sources, or other foreground components not smooth in the angular directions, can oscillate in and out of a side-lobe’s maxima. This can add some structure to the otherwise smooth foreground spectra and degrade the efficiency of the foreground clean. If this causes an increase in the foreground residuals in the cleaned \hi IM, then these could plausibly be non-Gaussian and bias the bispectrum. We simulated this effect in the simulations (see \secref{sec:NonGaussBeam} for details) and show the bispectrum measurements, also in \autoref{fig:CosineBeamModel}. We did not include polarisation leakage effects from the foregrounds since this made signal-to-noise very poor with very large errors, 
however we were still able to model the combined effects of a frequency-dependent beam plus polarised foregrounds. The foreground cleaned results in \autoref{fig:CosineBeamModel} (red-squares) used a PCA clean with $N_\text{fg}=4$ and $\kFG=5\times 10^{-3}\,h/\text{Mpc}$ for the modelling. It is encouraging to note that this is still not causing drastic effects to the success of the model as shown by the $\chi^2_\text{dof}$ (displayed in the legend). We stress that the simulated beam pattern we have used is a simplification, designed to investigate some of the key factors we felt could impact the bispectrum. For example, we are still assuming the beam is symmetrical, which may not be the case for a radio telescope array like MeerKAT \citep{2019arXiv190407155A}. Furthermore, the cosine beam pattern is a generalised approximation, but the precise beam pattern will be unique to each instrument. However, given the little impact we have seen from our results in \autoref{fig:CosineBeamModel}, it appears unlikely that these subtle differences will significantly alter results.

Some improvements could be made to both cosine beam results by performing a re-smoothing of the data to a common resolution. Alternatively, improvements to the beam bispectrum modelling could be made by accounting for a frequency-dependent beam size and the side-lobes in the beam pattern. However, this is beyond the scope of this work. A further general conclusion is that our bispectrum measurements appear to largely avoid biased effects concentrated to the scales relevant to the ripple frequency ($T$) in the frequency-dependent beam size (\autoref{eq:ripple}). These effects were seen in \citet{Matshawule:2020fjz}, in their radial 1D power spectra measurements. We believe that we do not see these effects because when measuring the spherically averaged bispectrum this ripple in the beam, which manifests on precise radial modes related to the $T=20\,\text{MHz}$ frequency, is mostly spread out amongst 3D modes where $k^2=k_{\!\myparallel}^2 + k_\perp^2$. Whereas in the \citet{Matshawule:2020fjz} results, this caused a very concentrated effect at scales related to $k_{\!\myparallel}\sim 2\pi/T$, in the $P_\text{1D}(k_{\!\myparallel})$ radial power spectrum.

\section{Conclusion}\label{sec:Conclusion}

In this paper we have demonstrated how the main observational effects relevant to a \hi IM survey can be modelled in a bispectrum analysis. We have provided validation tests for these models through comparisons with measured bispectra of simulated \hi IM data from an $N$-body semi-analytical technique for the cosmological \hinospace, including emulated effects from the telescope beam and a foreground clean, as well as redshift space distortions. This modelling framework was developed in the context of a low-redshift, single-dish IM survey, however, these models could be transferable or extended for low-$z$ interferometer \hi IM surveys as well as EoR surveys.\newline
\\
\noindent The main conclusions we draw from this work are:

\begin{itemize}[leftmargin=*]

\item Using second order perturbation theory and closely following previous work for optical galaxy surveys, we find that the \hi IM bispectrum can be modelled well at scales $k\lesssim 0.15\,h/\text{Mpc}$. As expected, RSD introduce a noticeable effect into the \hi bispectrum (\autoref{fig:NoRSD}) which can be modelled with standard redshift space bispectrum kernels (\autoref{eq:Kaiser} and \autoref{eq:Z2}). Including a FoG term in this RSD modelling we found that a non-zero value for $\sigma_\text{B}$ made agreement between data and model worse. This is due to the failing of the model at high-$k$ scales that cannot be treated perturbatively. However, since these scales are greatly affected by beam effects in \hi IM, we concluded that this model was sufficient for our purposes. We omitted non-local bias corrections in our modelling, since the primary aim of this work is investigating the more dominant IM observational effects. However, these would be required for precise parameter estimation.
\\
\item Applying a Gaussian beam to our simulations, with size defined by physical scale $R_\text{b}$, we derived a model for the effects on the measured bispectrum from these smoothed fields. We found the beam produced a trivial damping to high-$k$ in the equilateral bispectrum (\autoref{fig:BeamModel}) but in the isosceles cases, there were some more complex distortions (\autoref{fig:Beamisos}). In summary, the beam model performed well in all cases and was able to match the distortions caused from the beam.
\\
\item By adding simulated foreground contamination to our \hi IM data, then cleaning these using PCA, we were able to study the effects of foreground cleaning on the \hi IM bispectrum. As expected we found this mainly damped small-$k$ modes but crucially we were able to apply a phenomenological model to the bispectrum that agreed well with the foreground cleaned results. This technique relies on tuning a single free parameter ($\kFG$), which governs how strong the damping from foreground cleaning is.
\\
\item We used a reduced $\chi^2_\text{dof}$ test throughout to measure the goodness-of-fit between our models and simulations and in general found good agreement with $\chi^2_\text{dof}\sim 1$ for RSD, beam, and foreground models. 
\\
\item 
Since the thermal noise from the radio telescope should be Gaussian (white noise), the \hi IM bispectrum should be immune to thermal noise bias. We demonstrated how only unrealistically high levels of noise cause noticeable reduction to the $S/N$ (\autoref{fig:SNR} - top-panel). However, this is under the assumption that contributions from other systematics can be controlled at an exquisite level, which is a major ongoing challenge for \hi IM.
\\
\item We relaxed our assumption of a perfectly Gaussian, frequency-independent beam to investigate the impact this would have on bispectrum modelling. We used a cosine beam pattern with a frequency dependence and whilst agreement by-eye is still good, we found that this does increase the $\chi^2_\text{dof}$ (\autoref{fig:CosineBeamModel} - blue-triangles). However, improvements to this should be possible either by attempting to model the non-Gaussianity in the beam, or treating the actual data by re-convolution to a common Gaussian beam resolution as is typically done in \hi IM data surveys.
\\
\item We also included full-sky foregrounds convolved with the more realistic cosine beam, to investigate the effect shifting side-lobes will have on foreground contamination on a bispectrum measurement. We found no clear evidence that this increases modelling problems (\autoref{fig:CosineBeamModel} - red-squares), however, this may not be the case if we were just probing radial modes along the line-of-sight as identified in \citet{Matshawule:2020fjz} in the radial 1D power spectrum.
\\
\item We examined the impact on the bispectrum's $S/N$ from foregrounds and the beam separately in \autoref{fig:SNR} (bottom-panel). This showed that the beam has a significantly larger impact decreasing the overall $S/N$ by $61.9\%$ relative to the foregrounds which, even with polarisation leakage, only decreases the $S/N$ by $7.9\%$. This is mainly due to the fact that the beam damps modes with high $S/N$, whereas foregrounds mainly damp low-$k$ modes. However, in a realistic situation it is very likely that foreground removal effects will be much harder to  to characterise (and model) than the beam, which means they can pose a much bigger challenge for precise and accurate parameter estimation (e.g. BAO and the growth of structure  \citep{Soares:2020zaq}, or primordial non-gaussianity \citep{Cunnington:2020wdu}).

\end{itemize} 

\noindent In future work it would be interesting to further explore non-linear effects on the \hi bispectrum. This would require higher resolution simulations, ideally with hydrodynamics to allow analysis of the particularly complex distribution of \hi on smaller scales. It would be revealing to see whether non-linear effects on higher-order statistics, such as the bispectrum, significantly differ between \hi IM and optical galaxy surveys. Furthermore, our work could be extended to include parameter estimation to see if the bispectrum including the observational effects can help constrain e.g. the \hi bias. However, we reiterate that this would require the extension of our theoretical modelling to include a non-local bias correction, something we omit in this work. Indeed, including a more robust modelling of non-linear scales is likely to be particularly relevant for interferometers with better resolution than single-dish experiments. Exploiting the non-linear scales with interferometers can assist in breaking degeneracies and improve parameter constraints \citep{Castorina:2019zho}. Lastly, including higher order multipoles in this analysis would extend upon \citep{Cunnington:2020mnn} which studied IM observational effects in the power spectrum multipoles. IM observational effects should also leak signal into higher-$\ell$ thus it would be revealing to see if including higher-order multipoles improves parameter constraints for the \hi IM bispectrum, as has been shown to be the case for the power spectrum \citep{Soares:2020zaq}.

\section*{Acknowledgements}

We thank Bernhard Vos Gin\'{e}s and Santiago Avila for their help regarding the cold gas mass simulations. We also thank Paula Soares and Marta Spinelli for useful discussions and feedback. SC is supported by STFC grant ST/S000437/1. CW's research for this project was supported by a UK Research and Innovation Future Leaders Fellowship, grant MR/S016066/1. AP is a UK Research and Innovation Future Leaders Fellow, grant MR/S016066/1, and also acknowledges support by STFC grant ST/S000437/1. This research utilised Queen Mary's Apocrita HPC facility, supported by QMUL Research-IT \url{http://doi.org/10.5281/zenodo.438045}. We acknowledge the use of open source software \citep{scipy:2001,Hunter:2007,  mckinney-proc-scipy-2010, numpy:2011,  Lewis:1999bs}.

\section*{Data Availability}

The data underlying this article will be shared on reasonable request to the corresponding author.




\bibliographystyle{mnras}
\bibliography{Bib} 




\appendix

\section{Simulated Data}

\subsection{Cosmological \hi}\label{app:CosmoHI}

To generate our \hi cosmological signal we used the \textsc{MultiDark-Galaxies} $N$-body simulation data \citep{Knebe:2017eei} and the catalogue produced from the \textsc{SAGE} \citep{Croton:2016etl} semi-analytical model application. These galaxies were produced from the dark matter cosmological simulation \textsc{MultiDark-Planck} (MDPL2) \citep{Klypin:2014kpa}, which follows the evolution of 3840$^3$ particles in a cubical volume of $1\,(\text{Gpc}/h)^3$ with mass resolution of $1.51\times10^9h^{-1}$M$_\odot$ per dark matter particle. The cosmology adopted for this simulation is based on \textsc{Planck}15 cosmological parameters \citep{Ade:2015xua}, with $\Omega_\text{m} = 0.307$, $\Omega_\text{b} = 0.048$, $\Omega_\Lambda = 0.693$, $\sigma_8 = 0.823$, $n_\text{s} = 0.96$ and Hubble parameter $h=0.678$. The catalogues
are split into 126 snapshots between redshifts $z= 17$ and $z=0$. In this work we chose low-redshift, post-reionisation data to test our models and use the snapshot at $z=0.39$ to emulate a MeerKAT-like survey performed in the L-band ($899 < \nu < 1184\,\text{MHz}$, or equivalently $0.2<z<0.58$). We obtained this publicly available data from the Skies \& Universes web page\footnote{\href{http://www.skiesanduniverses.org/page/page-3/page-22/}{www.skiesanduniverses.org}}. 

We used each galaxies (x, y and z) coordinates and placed them onto a grid with $n_\text{x}, n_\text{y}, n_\text{z} = 256,256,256$ pixels and $1\,(\text{Gpc}/h)^3$ in physical size. To simulate observations in redshift space inclusive of RSD, we utilised the peculiar velocities of the galaxies. Assuming the LoS is along the z-dimension and given the plane-parallel approximation is exact for this Cartesian data, RSD can be simulated by displacing each galaxy's position to a new coordinate $z_\text{RSD}$ given by
\begin{equation}\label{PlaneParaEq}
    \text{z}_\text{RSD} = \text{z} + \frac{1+z}{H(z)}h\, v_{\!\myparallel} \, ,
\end{equation}
where $v_{\!\myparallel}$ is the galaxy's peculiar velocity along the LoS (z-dimension) which is given as an output of the simulation in units of $\text{km}\,\text{s}^{-1}$.

To simulate the contribution to the signal from each galaxy, we used the cold gas mass $M_\text{cgm}$ output from the \multidark\ data and from this we can infer a \hi mass with $M_\hinospace = f_\text{H} M_\text{cgm}(1-f_\text{mol})$ where $f_\text{H}=0.75$ represents the fraction of hydrogen present in the cold gas mass and the molecular fraction is given by $f_\text{mol}=R_\text{mol}/(R_\text{mol}+1)$ \citep{Blitz:2006nc}, with $R_\text{mol} \equiv M_{H_2}/M_\hinospace = 0.4$ \citep{Zoldan17}. It is this \hi mass that we binned into each voxel with position $\boldsymbol{x}$, to generate a data cube of \hi masses $M_\hinospace(\boldsymbol{x})$,  which should trace the underlying matter density generated by the catalogue's $N$-body simulation for the snapshot redshift $z$. These \hi masses are converted into a \hi brightness  temperature for a frequency width of $\deltadiff \nu$ subtending a solid angle $\deltadiff \Omega$ given by
\begin{equation}\label{THIequation}
    T_\hinospace(\boldsymbol{x},z) = \frac{3h_\text{P}c^2A_{12}}{32\pi m_\text{h}k_\text{B}\nu_{21}}\frac{1}{\left[(1+z)r(z)\right]^2}\frac{M_\hinospace(\boldsymbol{x})}{\deltadiff \nu \, \deltadiff \Omega} \, ,
\end{equation}
where $h_\text{P}$ is the Planck constant, $A_{12}$ the Einstein coefficient that quantifies the rate of spontaneous photon emission by the hydrogen atom, $m_\text{h}$ is the mass of the hydrogen atom, $k_\text{B}$ is Boltzmann's constant, $\nu_{21}$ the rest frequency of the 21cm emission and $r(z)$ is the comoving distance out to redshift $z$ (we will assume a flat universe). Since \hi simulations on this scale have a finite halo-mass resolution, there will be some contribution from the \hi within the lowest-mass host haloes which is not included in the final $T_\hinospace$ signal. To account for this, it is typical for a rescaling of the final $T_\hinospace$ to be performed to bring the mean \hi temperature, $\overline{T}_\hinospace$, in agreement with the modest data constraints we have for this value. For the effective redshift of our data, $z=0.39$, we used a fiducial value of $\overline{T}_\hinospace=0.0743\,\text{mK}$ which our maps were re-scaled to. Lastly, \hi  data is in general unavoidably mean-centred due to foreground cleaning processes, and therefore the final calibrated data is in the form of a temperature fluctuation given by
\begin{equation}\label{eq:deltaT}
	\deltadiff T_\hinospace(\boldsymbol{x},z) = T_\hinospace(\boldsymbol{x},z) - \overline{T}\hspace{-0.5mm}_\hinospace(z) \, .
\end{equation}
This represents the final form of our simulated data and examples of these were shown in \autoref{fig:Maps}.

\subsection{21cm Foreground Simulations}\label{app:FGSims}

The observed IM data can be approximately decomposed as $T_\text{obs} = T_\hinospace + T_\text{fg}$. To produce the $T_\text{fg}$ component we simulated different foreground processes, including galactic synchrotron, free-free emission and point sources. We also included the effects of polarisation leakage which will act as an extra component of foreground with non-smooth spectra, thus posing an increased challenge for the foreground clean. The foregrounds we used can thus be decomposed as $T_\text{fg} = T_\text{sync} + T_\text{free} + T_\text{point} + T_\text{pol}$, which represent the synchrotron, free-free, point sources and polarisation leakage.

We briefly summarise the simulation technique for these components but for a full outline we refer the reader to \citet{Cunnington:2020njn} and \citet{Carucci:2020enz} where they were also used. Furthermore, a full-sky realisation is openly available from \citet{FGsim}. The synchrotron emission is based on Planck Legacy Archive\footnote{\href{http://pla.esac.esa.int/pla}{pla.esac.esa.int/pla}} FFP10 simulations of synchrotron emission at $217$ and $353\,\text{GHz}$ formed from the source-subtracted and destriped $0.408\,\text{GHz}$ map. The free-free simulation is from the FFP10 217\,GHz free-free simulation at which is a composite of the \citet{cliveff} free-free template and the WMAP MEM free-free templates. The point sources are based on the empirical model of \citet{batps} and makes the assumption that point sources over $10\,\text{mJy}$ will be identifiable and thus can be removed. Lastly, we simulated polarisation leakage with the use of the \texttt{CRIME}\footnote{\href{http://intensitymapping.physics.ox.ac.uk/CRIME.html}{intensitymapping.physics.ox.ac.uk/CRIME.html}} software \citep{crime}, which provides maps of Stokes Q emission at each frequency and we fix the polarization leakage to $0.5\%$ of the Stokes Q signal.

For the foregrounds we assumed they have been observed in a frequency range of $900 < \nu <1156\,\text{MHz}$, consistent with the $z=0.39$ redshift for the cosmological simulation. Each of the 256 map slices along the z-direction acts as an observation in a frequency channel giving a channel width of $\deltadiff \nu = 1\,\text{MHz}$. This therefore emulates the spectral distinction between the cosmological \hi and foregrounds utilised in the foreground clean. From the full-sky foreground map we cut a region of sky centred on the Stripe82 region of sky, a field well observed by surveys. The size of this sky region is $54.1 \times 54.1\,\text{deg}^2$ which corresponds to the size of a $1\,(\text{Gpc}/h)^2$ patch at the $z=0.39$ snapshot redshift of our cosmological simulation. A map of the foreground signal was shown in \autoref{fig:Maps} (top-right).

\subsection{Full-Sky Beam Convolution}\label{app:BeamSims}

In the more complex pattern of a cosine beam, there exist a number of side-lobes as shown by \autoref{fig:BeamPattern}. Since the side-lobes can continue out to very wide distances from the central pointing ($\theta=0$), and potentially pick-up dominant signal from stronger regions of the sky e.g. the galactic plane, just using this beam pattern on our $1\,(\text{Gpc}/h)^2$ patch of sky will not sufficiently emulate this behaviour. Instead, we carry out a full-sky convolution of the foregrounds which should produce any of the effects we discussed in our targeted region. We then cut this sky region and overlay our \hi simulated data. Whilst this approach is not entirely consistent with a real experiment, it is sufficient for our purposes for investigating these severe observational effects on a bispectrum measurement. To carry out the full-sky convolution we decomposed the map into spherical harmonics $Y_{\ell m}$ such that
\begin{equation}
    T_\text{fg}(\nu, \boldsymbol{\theta}) = \sum_{\ell=0}^{\infty} \sum_{m=-\ell}^{\ell} a_{\ell m}(v) Y_{\ell m}(\boldsymbol{\theta})\,,
\end{equation}
where the harmonic coefficients $a_{\ell m}$ describe the amplitudes of the fluctuations in spherical harmonic space. This allows the convolution to be applied as a simple product in spherical harmonic space between the harmonic coefficients $a_{\ell m}$ and the harmonic coefficients for the cosine beam function we are using, such that the new convolved coefficients, with correct normalisation factors, are given by
\begin{equation}
    \tilde{a}_{\ell m}(\nu)=\sqrt{\frac{4 \pi}{2 \ell+1}} a_{\ell m}(\nu) \frac{b_{\ell}(\nu)}{\sqrt{4 \pi} b_{0}(\nu)}\,,
\end{equation}
where $b_\ell$ are the beam harmonic coefficients, which assuming a symmetrical beam function can be given by
\begin{equation}
    b_{\ell}(v)=\int \mathcal{B}_{\mathrm{C}}(\nu,\boldsymbol{\theta})\,Y_{\ell 0}^{*}(\boldsymbol{\theta}) \,\text{d}\boldsymbol{\theta}\,.
\end{equation}
where $\mathcal{B}_{\mathrm{C}}$ is given by \autoref{eq:CosineBeam}. We refer the reader to \citet{Matshawule:2020fjz} for a more focused investigation into beam effects on the efficiency of foreground cleaning, than what we intend to carry out here. In this case of simulating a more realistic beam, we assumed a dish size of $D_\text{dish}=13.5\,\text{m}$ and the frequency range as before of $900<\nu<1156\,\text{MHz}$ (with $\delta\nu=1\,\text{MHz}$), consistent with a MeerKAT L-band survey.

\section{Observational Effects on the Bispectrum Covariance}\label{app:Covariance}

\begin{figure}
    \centering
    \includegraphics[width=0.95\textwidth]{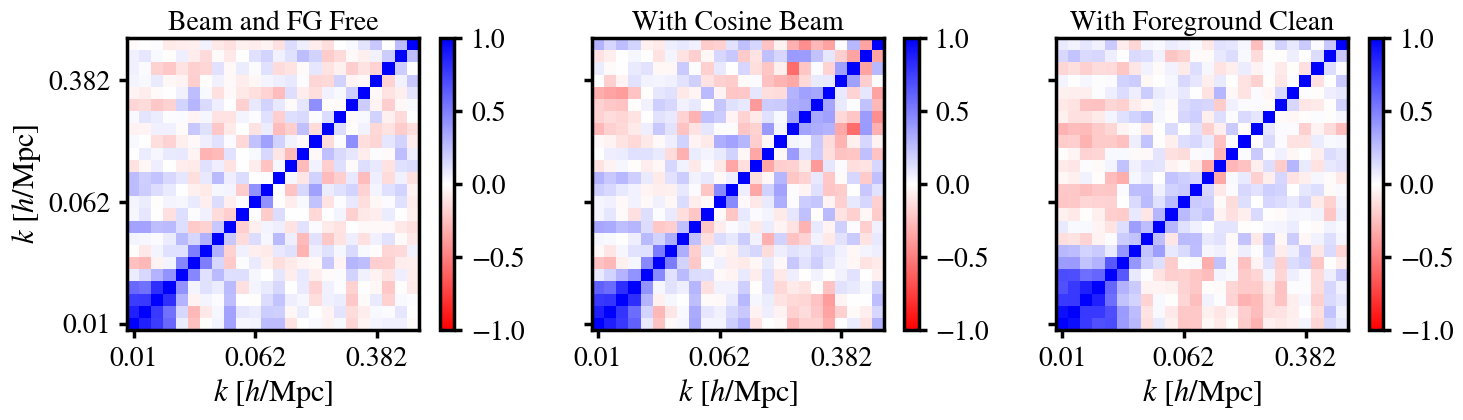}
    \caption{Correlation between bins for the equilateral \hi IM bispectrum monopole. \textit{Left}-panel: without any effects from a telescope beam or foregrounds. The other panels demonstrate the low impact on covariance from observational effects caused by the beam (\textit{centre}) and foreground cleaning (\textit{right}).}
    \label{fig:Covariance}
\end{figure}

The covariance matrix is used to estimate the errors from the bispectrum estimator and indicates whether different $k$ bins are correlated. In optical galaxy surveys, it has been found that there is more correlation between bins in the bispectrum compared to the power spectrum \citep{Gil-Marin:2016wya}. In this work we did not aim to investigate this in detail, since a robust covariance estimation for the bispectrum is complex due to high number of triangle bins and typically requires a large number of mocks \citep{Byun:2020rgl}. Instead, we used a jackknife routine, splitting our data into $N_\text{jack}=64$ sub-samples and estimating the covariance with \citep{Norberg:2008tg} 
\begin{equation}
    C\left(k_{i}, k_{j}\right)=\frac{(N_\text{jack}-1)}{N_\text{jack}} \sum_{n=1}^{N_\text{jack}}(k_{i}^{n}-\bar{k}_{i}) \, (k_{j}^{n}-\bar{k}_{j})\,,
\end{equation}
where $\bar{k}_{i,j}$ are the mean averages over the $N_\text{jack}$ measurements. This approach allows a simple, yet sufficient means to check correlations between bins (i.e., check the assumption that the covariance is diagonal), infer error-bars, and also study if the covariance is affected by the observational effects from the beam or foreground cleaning. We show the correlation matrix defined as $R_{ij} = C_{ij}/\sqrt{C_{ii}C_{jj}}$ between $k$ bins in \autoref{fig:Covariance}. We show this only for the equilateral configuration for simplicity. The left-panel shows the \hi IM without a beam or foreground contamination. We can see that the diagonal covariance assumption is reasonable, except perhaps at low-$k$ where bins become more correlated, likely due to our logarithmic binning scheme. More importantly for this work, we find there is little impact from a frequency-varying telescope beam with side lobes (centre-panel), or from data with polarised foregrounds cleaned using a $N_\text{fg}=10$ PCA method (right-panel). There is a slight increase in correlation at high-$k$ for the beam case, and at small-$k$ for the foreground case, as one may expect \citep{Wolz:2013wna}. Therefore, this preliminary study suggests that neither the beam or foregrounds should cause large problems for the bispectrum covariance.

\section{Aliasing Corrections to the Bispectrum}\label{app:Aliasing}

Our \hi IM bispectrum measurements do not currently include any corrections for potential aliasing effects, which could in principle be causing some discrepancies at high-$k$. Indeed, scales where $k>2k_\text{Nyq}/3$ (where $k_\text{Nyq}$ is the Nyquist frequency) are likely to become biased in the bispectrum due to aliasing contributions \citep{Sefusatti:2015aex}, which for our simulations equates to $k\sim0.54\,h/\text{Mpc}$. We ran tests on a higher resolution gridding with $512^3$ cells and found results follow this prediction with a noticeable bias beginning to form at $k>2k_\text{Nyq}/3$ in the case of an equilateral triangle configuration. However, below this at $k<2k_\text{Nyq}/3$, results between the $256^3$ and the $512^3$ gridding scheme were consistent. 

A simplified approach to correct for the effects of mass assignment can be made to the density field whereby $\delta^\text{true}(\boldsymbol{k})\approx\delta^\text{meas}(\boldsymbol{k})/W(\boldsymbol{k})$, where $W$ is the Fourier transform of the mass assignment function \citep{Jing:2004fq}. Indeed this approach has been used in real data bispectrum analysis on optical galaxy redshift surveys \citep{Gil-Marin:2014sta}. However, the more thorough approach would need to replicate that typically used in aliasing corrections to the power spectrum, which involves an iterative process and requires a priori knowledge of the target power spectrum one expects to measure. This is less trivial to calculate in the case of the bispectrum where triangle-shape dependencies would cause issues. This is discussed in \citet{Sefusatti:2015aex}, which also proposes corrections using interlacing techniques performed directly to density field in Fourier space. These would therefore be naturally applicable to higher order measurement of this density field such as the bispectrum. However, since we found no noticeable effects on the scales we were interested in, we did not investigate such corrections in the context of \hinospace, but encourage future work into this, since it should become necessary if aiming to contribute to precision cosmology.


\bsp	
\label{lastpage}
\end{document}